\documentclass[usegraphicx,usedcolumn,usenatbib]{mn2e}
\usepackage{color,epsfig,rotating,amssymb,longtable}
\usepackage{array,colortbl}
\usepackage{lscape}
\usepackage[colorlinks=true,linkcolor=magenta,citecolor=blue,breaklinks=true]{hyperref}
\usepackage{longtable}
\usepackage{ifpdf}

\ifpdf
\usepackage{epstopdf}
\else
\usepackage[dvips]{graphicx}
\usepackage{lscape}
\fi
\def\ergs{~erg\,s$^{-1}$}
\def\ergsc{~erg\,s$^{-1}$cm$^{-2}$}
\def\kms{~km\,s$^{-1}$}

\def\HI{H{\sc i}}
\def\HII{H{\sc ii}}

\def\Haro{Haro\,15}

\usepackage{color}

\newcommand\ion[2]{#1~{\sc {#2}}\relax} 

\definecolor{dgreen}{rgb}{0,.5,.1} 
\definecolor{pink}{rgb}{.9,.4,.7}

\usepackage{graphicx}

\title[Haro 15: Internal kinematic]{High resolution spectroscopy of the BCD galaxy Haro\,15: I. Internal kinematics} 
\author[V. Firpo et al.]{Ver\'onica Firpo$^{1}$\thanks{vfirpo@fcaglp.unlp.edu.ar}, 
Guillermo Bosch$^{1}$\thanks{IALP-CONICET, Argentina.}, Guillermo F. H\"agele$^{1,2}$\thanks{CONICET, Argentina.}
\newauthor
 \'Angeles I. D\'{\i}az$^{2}$ and Nidia Morrell$^{3}$ \\  
$^{1}$ Facultad de Ciencias Astron\'omicas y Geof\'{\i}sicas, Universidad Nacional  
de la La Plata, Paseo del Bosque s/n, 1900 La Plata, Argentina.\\ 
$^{2}$ Departamento de F\'{\i}sica Te\'orica, C-XI, Univerdidad Aut\'onoma de
Madrid, 28049 Madrid, Spain.\\
$^{3}$ Las Campanas Observatory, Carnegie Observatories, Casilla 601, La Serena, Chile.} 

\begin{document} 

\date{Accepted . Received ; in original form } 


\maketitle

\begin{abstract} 
{
Using echelle spectroscopy, obtained at Las
Campanas Observatory, we present a detailed study of the internal kinematics of the nebular material in multiple knots of the blue compact dwarf galaxy \Haro. 
A detailed analysis of the complex emission line profiles show the presence of an underlying broad component in almost all knots, and the brightest star-forming region shows unmistakable signs for the presence of two distinct narrow kinematical components.
We also study the information that our analysis provides regarding the motion of the individual knots in the \Haro\ galaxy potential, confirming that they follow galactic rotation. Finally, we examine the relation between their velocity dispersion and luminosity, finding that almost all knots follow the relation for virialised systems. This holds for the strong narrow components identified in complex fits and for single profile fits, although the latter show a flatter slope. In agreement with previous findings, in this paper we show that the existence of multiple kinematical components among massive starbursts cannot be overlooked, as it has a noticeable effect on any subsequent analysis that relies on basic parameters.}
\end{abstract}


\section{Introduction}

Already in the original list of blue galaxies with emission lines by Haro (1956) who described this object as ``minute cometary nebula'', \Haro\ has later been included in several compilations of Blue Compact Galaxies (BCG).  Several studied at all frequencies have been carried out over this object.  In particular, the optical spectroscopy was analysed by \cite{1985AJ.....90.1457H}; \cite{1991AJ....101.2034M}; \cite{2002A&A...396..503K}; \cite{2005A&A...437..849S}, between others. Owing to the detection of the \ion{He}{ii}~$\lambda$4686 emission line by \cite{1999IAUS..193..604K}, \cite{1999A&AS..136...35S} has classified this galaxy as a Wolf-Rayet (WR) galaxy.  \cite{2010A&A...516A.104L} also found a blue WR bump supporting the WR nature of \Haro.
The H$\alpha$ image shown in \cite{2001ApJS..133..321C} shows a knotty morphology with the starburst region resolved in a large number of clumps which appear scattered over the entire galaxy. 
The integrated H$\alpha$ luminosity yields a star formation rate (SFR) of 3.3 M$_{\odot}$ yr$^{-1}$ \citep{2008A&A...491..131L}. 
\cite{2010A&A...521A..63L} analysed the SFR in \Haro\ using multi-wavelength data. The derived values using H$\alpha$, FIR (far-infrared) and radio data are very similar, but the value derived using the FUV (far-ultraviolet) luminosity is twice the others, confirming that a young stellar population is dominating the light of the galaxy and suggesting that the starburst phenomenon in \Haro\ started some time ago (at least, 100 Myr ago). From the morphological point of view, \Haro\ has been classified as an (R)SB0 peculiar galaxy by \cite{RC3.9} and in fact, in the deep images taken with the 2.2m CAHA telescope shown in \cite{2008A&A...491..131L}, the spiral morphology of the galaxy can easily be appreciated. Two high surface brightness concentrations can be distinguished, named A and C by the authors. Both show blue colours and high FUV emission, indicative of recent and on-going star formation activity, further supported by the presence of the WR features. 

Long slit optical spectroscopy of \Haro\ has been presented in \cite{2009A&A...508..615L} covering four main regions in the galaxy: the center (named C by the authors) and three bright regions located ESE (named A), WNW (named D) and NE (named B). The spectrum of the central region shows strong nebular emission lines and prominent stellar absorption wings in the \HI\ Balmer lines, evidence of an underlying stellar population which the authors estimate to be around 500 Myr old. Only emission lines are seen in regions A and B and the spectrum of region D results too noisy to be analysed. These authors derived an oxygen abundance of 12+log(O/H)= 8.37$\pm$0.10 and 8.10$\pm$0.06 for regions A and B, respectively. Based on these data the authors conclude that all the observed knots can be classified as typical \HII\ regions. 

\Haro\ has an absolute magnitude M$_{B}$ = -20.69, a surface brightness $\mu_{B}$= 18.56 mag arcsec$ ^{-2} $ and a colour B-V= 0.33 \citep{2001ApJS..133..321C}. At a distance of 86.6 Mpc \citep{RC3.9}, Haro 15 meets the criteria for a Luminous Compact Blue Galaxy (LCBG) \citep{2004AJ....128.1541H} despite the fact of its brightness distribution showing an exponential profile with a scale length of 1.37 kpc.
The nature of this type of galaxies is not yet clear since they probably constitute a mixed population of starburst galaxies. Some authors suggest that LCBGs might represent the final outcome of a merger between a dwarf elliptical and a gas rich dwarf galaxy or \HI\ cloud \citep{Ostlin98,2008A&A...479..725C}. The interactions taking place during the merging process would act as the starburst trigger \citep[see e.g.][]{2004A&A...428..425L,2006A&A...449..997L}.
The images of \Haro\ showing two separated nuclei surrounded by a more regular, roughly elliptical envelope with twisted isophotes, together with different features and faint extensions as reported by \cite{2001ApJS..133..321C} give support to this picture. Other galaxy properties like different kinematics and chemical abundances between regions A and B may also indicate that \Haro\, is probably experiencing a minor merger \citep{2010A&A...521A..63L}. On the other hand, a reliable determination of the mass of this kind of objects is needed in order to decide on the evolutionary path they may follow: becoming dwarf spheroidal galaxies \citep{2007AJ....134.2455H} or the bulges of small spirals \citep{2001ApJ...550..570H}. However, interactions and mergers, as well as feedback processes result of intense star formation, might reflect in a peculiar gas kinematics that prevent the derivation of their dynamical masses from the gas velocity widths.  

Echelle spectroscopy provides a means to look for kinematically different components in the emission lines of the ionised gas since it reaches the spectral resolution needed to resolve the presence of structures within the emission-line profile, usually masked by its large supersonic width.
In \cite{2010MNRAS.406.1094F} we observed a residual present in the wings of several lines when fitting a single Gaussian profiles to the emission lines observed in the Giant \HII\ Regions within the galaxies NGC\,7479 and NGC\,6070. 
Basing on the variety studies that have been proposed in the literature to interpret the existence of the broad supersonic component measured in the emission line profile of Giant H{\sc ii} Regions, and whenever possible, we have evaluated the possible presence of a broad component \citep[][among
others]{MT96,M99,H07,2009MNRAS.396.2295H,Hagele+10} or two symmetric low-intensity components in the fit to the observed emission line profile widths \citep{CHK94,Relano05,2006A&A...455..539R}. In the present work, we have also found that all \Haro\ regions show evidence of wing broadening evident mainly in the H$\alpha$ line and confirmed in other
emission lines. 
Another common feature of Giant \HII\ Regions is the supersonic width of their emission line profiles, although the origin of these velocities is not yet clear \citep{TM81,R86,H86}. A relation between H$\alpha$
luminosity and velocity dispersion of the form L(H$\alpha$) $\propto \sigma^{4}$ is expected under the assumption of a gravitational origin of the ionised gas dynamics. Therefore, a study of the implications that the existence of several kinematically distinct components has on the L(H$\alpha$) vs. $\sigma$ relation is of great value.



In this paper we present echelle spectroscopy obtained with the 100-inch du Pont Telescope at Las Campanas
Observatory (LCO) of five different positions across \Haro\ which provide a velocity resolution of about 12 kms$^{-1}$ allowing the identification of different kinematical components of the gas and the measurement of their corresponding velocity dispersions. This study is part of a project to obtain high spectral resolution echelle data to determine the nature of Giant \HII\ Regions visible from the southern hemisphere and analyse the physical conditions of the ionised gas of these regions and Blue Compact Dwarf galaxies. In Section 2 we present the
observations and the data reduction. Section 3 presents the analysis of the emission-line profiles and discusses the results. Finally, the summary and conclusions of this work are given in Section 4.

\section{Observations and Data Reduction} 
\label{sec:observations} 


\begin{figure}
\begin{center}
\caption[Haro15 finding chart]{Finding chart for the Giant H{\sc ii} Region
  candidates observed in \Haro, identified by circles and labelled following \cite{2001ApJS..136..393C}. Wide Field Planetary Camera 2 H$\alpha$ image was obtained from the Multimission Archive at the Space Telescope Science Institute (MAST)}
\label{figHaro15chart}
\includegraphics[angle=0,width=.47\textwidth]{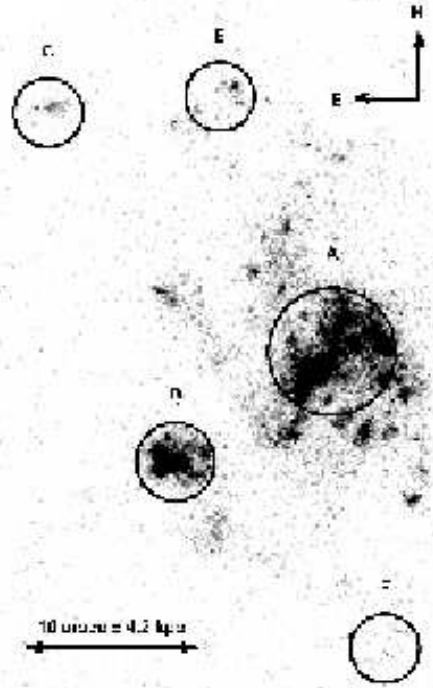}
\end{center}
\end{figure}

High resolution spectroscopy of \Haro\ was obtained using an echelle spectrograph attached to the 100-inch du Pont Telescope, Las Campanas Observatory (LCO), in July 19 and 20 of 2006. The spectral range covered by the observations was from 3400 to 10000~\AA. Observing conditions were good, with an average of 1 arcsec seeing and photometric nights. A 2$\times$2 binning was applied to the CCD in order to minimise the readout contribution to the final spectrum noise. With 1 arcsec effective slit width and 4 arcsec slit length, the spectral resolution achieved in our du Pont Echelle data was R$\simeq$25000: $\Delta\lambda$=0.25\AA\ at $\lambda$6000\AA, as measured from the FWHM of the ThAr comparison lines taken for wavelength calibration purposes. This translates into a velocity resolution of $\sim$12 kms$^{-1}$.\\
Considering the distance to \Haro\ and that the spatial resolution is limited by seeing, the smallest structure that can be resolved is 0.43 kpc ($\sim$ 1"$\sim$ 1.93 pixel)\\

Five different regions in \Haro\ were observed as shown in Figure \ref{figHaro15chart}. The spectra were obtaind as singles exposures of 1800 seconds. CALSPEC spectrophotometric standard star Feige\,110 \citep{Bohlin01} was also observed for flux calibration purposes with an exposure time of 1200 seconds. In addition, Th-Ar comparison spectra, milky flats (sky flats obtained with a diffuser, during the afternoon) and bias frames were taken every night. A journal of observations is shown in Table \ref{Regions}.

\begin{table}
\caption[Observed dates]{Journal of observations. The identifications in the first column correspond to the nomenclature used in this paper, following and extending that in \cite{2001ApJS..136..393C}; columns 2 and 3 list the observation date and the air mass respectively; columns 4 and 5 show the distance to the galactic center in arcsecs and kpc, respectivey; finally, column 5 lists the identification of the regions given in \cite{2008A&A...491..131L} referred to as LS notation. The exposure time for each region was 1800 seconds.}
\label{Regions}
\begin{center}
\begin{tabular}{@{}cccccc@{}}
\hline
Knots & Date &  sec z  & Dist.(") &  Dist.(kpc)&LS \\
\hline
\Haro\,A& 2006 Jul 19 &  1.2  & 0 & 0 &C\\
\Haro\,B& 2006 Jul 19 &  1.1  &  11 & 4.62 & A \\ 
\Haro\,C& 2006 Jul 19 &  1.1  & 20 & 8.4 & B  \\
\Haro\,E& 2006 Jul 20 &  1.2  & 16 &6.72 & \ldots  \\
\Haro\,F& 2006 Jul 20 &  1.1  & 17 &7.14  & \ldots  \\
\hline
\end{tabular}
\end{center}
\end{table}

The data analysis was carried out using the {\sc IRAF}\footnote{Image Reduction and Analysis Facility, distributed by NOAO, operated by AURA, Inc., under agreement with NSF.} software. After bias subtraction and flat field corrections by means of milky flats 
the bidimensional images were corrected for cosmic rays using the task {\sc cosmicrays} which detects and removes cosmic rays using a flux ratio algorithm. The corrected data were reduced by {\sc IRAF} routines following procedures similar to those described in \cite{Firpo05}. 

At the end of the process, we compared the red end of the wavelength calibrated spectra with the night-sky spectrum by \cite{O96}. This turned out to be a very reliable confirmation of the goodness of the wavelength solution, and we checked that differences between our wavelengths and the sky line wavelengths were below 0.04\AA.

Flux calibration was achieved by observing the CALSPEC spectrophotometric standard star, Feige\,110 (Bohlin et al.~\citeyear{Bohlin01}) whose flux was tabulated every 2\AA. Despite its relatively low brightness (V=11.83), Feige 110  is ideal for calibrating high resolution echelle spectra. The amount of defined intervals within an echelle order ranged from four to twelve, depending on the quality of the spectrum. The flux calibrations was performed as described in \cite{Firpo05}.

\section{Results and Discussion}
\label{sec:results and discussion} 

\subsection{Line profile analysis}
\label{sec:Line profiles analysis} 

We identified the hydrogen recombination lines, such as H$\alpha$ and H$\beta$, and collisionally excited lines, such as [\ion{N}{ii}]~$\lambda\lambda$6548,6584, [\ion{S}{ii}]~$\lambda\lambda$6717,6731 present in the spectra, making use of the known redshift, z=0.021371, for \Haro\ \citep{RC3.9}. The strong lines were used to analyse the structure of velocity profiles as they allow us to verify the existence of more than one component as described in \cite{2010MNRAS.406.1094F}. The adopted laboratory wavelengths were taken from the work of \cite{G-R05}. 

From the echelle calibrated spectrum, we cut the wavelength range where a given emission line is and we transformed from wavelength to velocity plane using the Doppler correction.  By measuring the central velocity (wavelength) and width of several emission lines we determine the radial velocities and velocity dispersions of the ionised gas in the differents star-forming regions of \Haro. The radial velocity and the intrinsic velocity dispersion ($\sigma_{\mathrm{int}}$), corrected for the instrumental and thermal contributions of each emission line, are also derived. For these observations we consider $\sigma_{i}$ = 5.2 kms$^{-1}$ as instrumental width. The thermal contribution was derived from the Boltzmann's equation ($\sigma_{t}$\,=\,2kT/m$_{a}$), where $k$ is the Boltzmann's constant, $T$ the kinetic temperature ($T \simeq 10^{4}K$) and $m_{a}$ the atomic mass of the corresponding element. Although a detailed determination of the electron density and temperatures, and chemical ionic and total abundances for each region will be presented in H\"agele et al.\ (2011, in prep., hereafter Paper~II), small changes of a few hundred degrees do not noticeably modify this correction, as previously discussed in \cite{2010MNRAS.406.1094F}. The fluxes were derived from the amplitude (A) and the FWHM  of the Gaussian profile obtained in the component fitting (F=1.0645$\times$A$\times$FWHM), and the corresponding errors were estimated taking into account the errors in these two parameters.

As already reported by Firpo and collaborators, in the present work we have also found that all \Haro\ emission knots show evidence of wing broadening, which is always found in the H$\alpha$ line profiles and it is usually also observed in the profiles of the bright emission lines. Making use of the iterative fitting of multiple Gaussian profiles we evaluated the presence of a broad component and more than one narrow component present in the emission line profile. In this case, we fitted a broad component, explaining the integral profile wings for all regions.

In the following subsections we will discuss our findings for each knot resulting from the profile fits.


\subsubsection*{\Haro\,A}

In our high resolution spectra \Haro\,A shows a complex structure which, although evident in radial velocity space, could not be spatially resolved.

\begin{figure*}
\begin{center}
\caption[A-strongFig]{Strongest emission lines from {\bf \Haro\ A} spectrum. Each panel includes flux calibrated spectrum, where the individual x-axis have been normalised to the observed radial velocity for comparison purposes. To enhance details at low luminosity levels, the y axes are shown in logarithmic scale. From top to bottom and left to right: H$\beta$, [\ion{O}{iii}]~$\lambda$5007, H$\alpha$, [\ion{N}{ii}]~$\lambda$6584 and [\ion{S}{ii}]~$\lambda$6717.
}\label{figA_broad}

\includegraphics[trim=0cm 0cm 0cm 0cm,clip,angle=0,width=8cm,height=5cm]{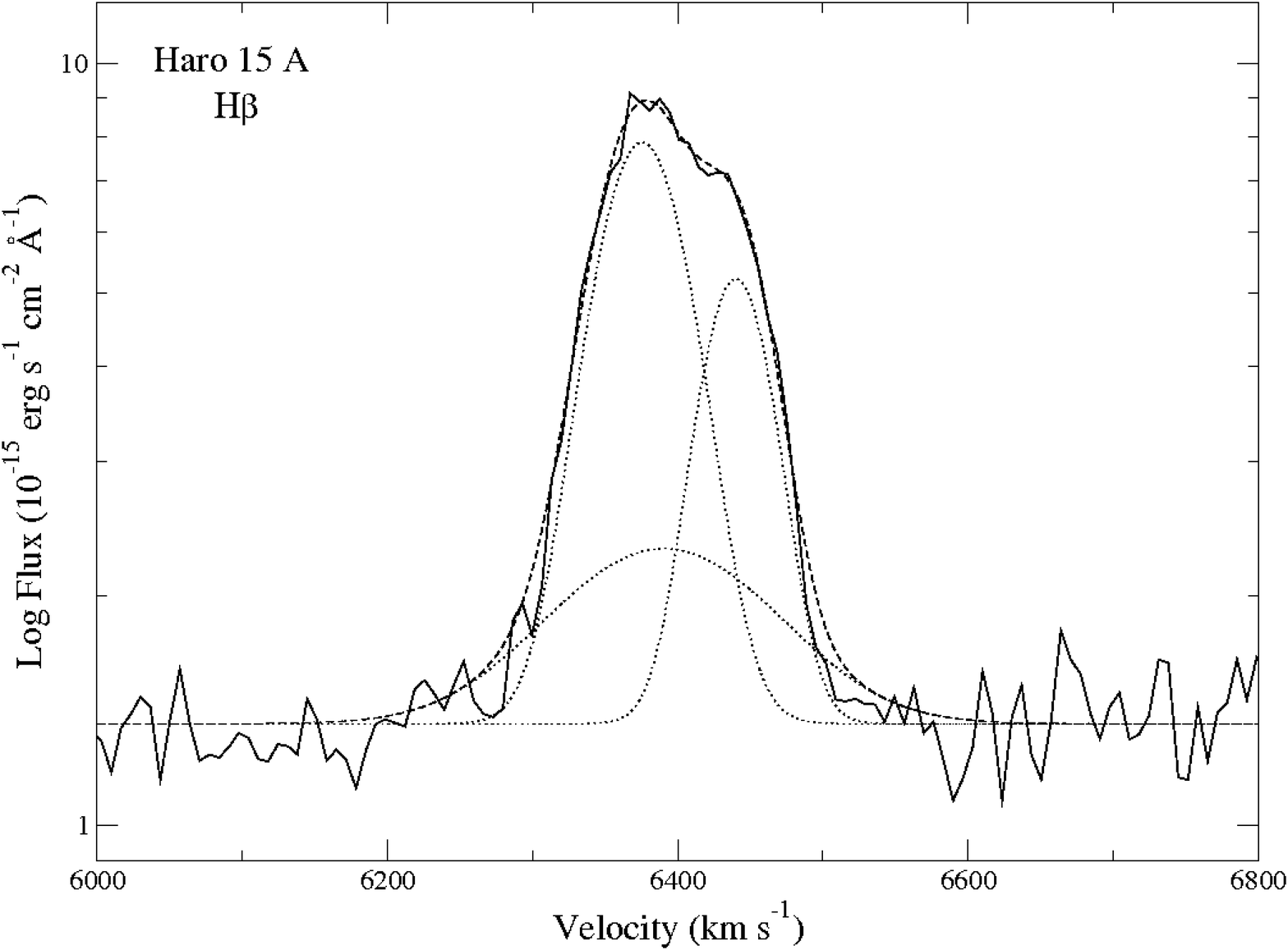}
\includegraphics[trim=0cm 0cm 0cm 0cm,clip,angle=0,width=8cm,height=5cm]{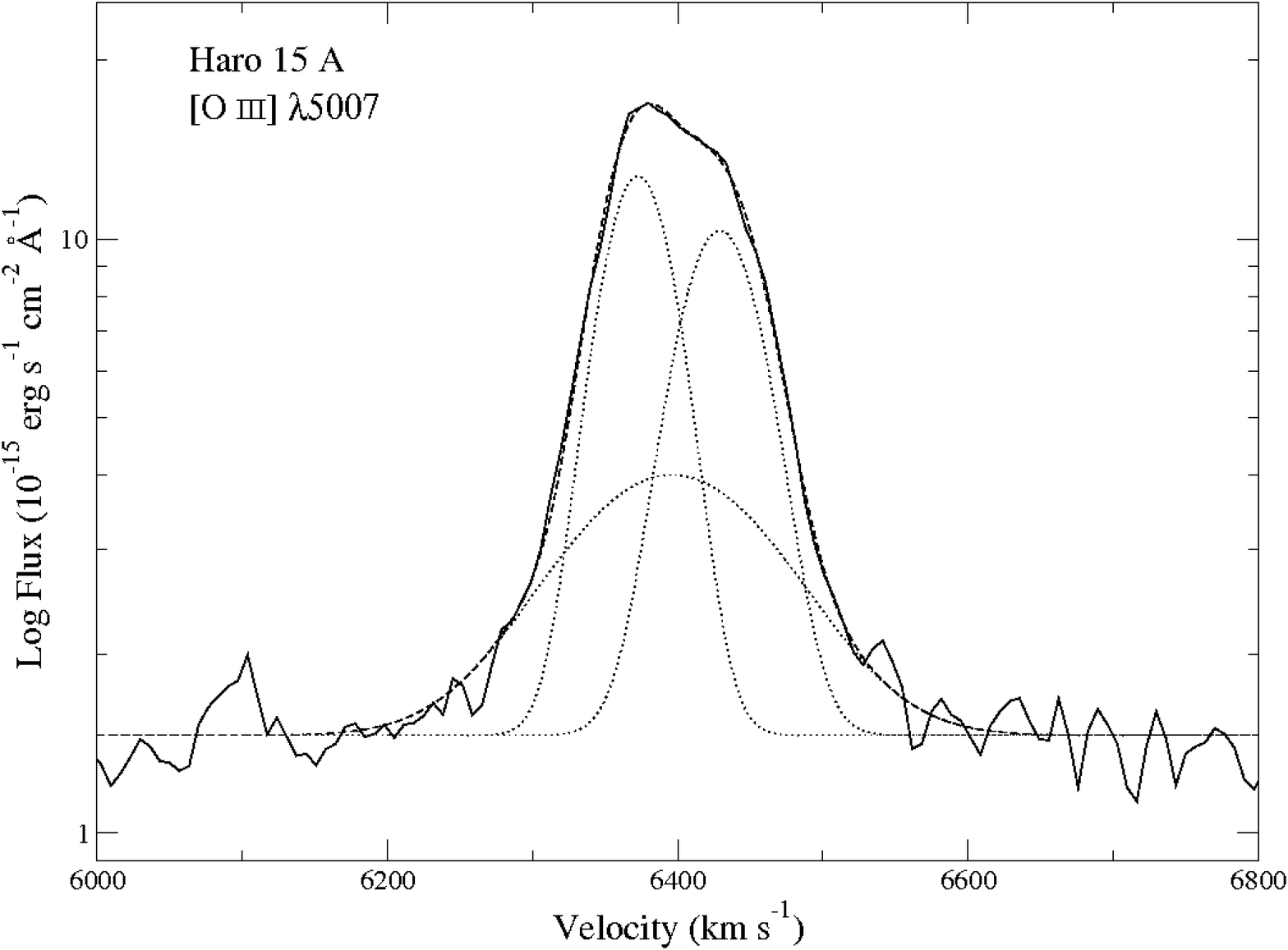}
\includegraphics[trim=0cm 0cm 0cm 0cm,clip,angle=0,width=8cm,height=5cm]{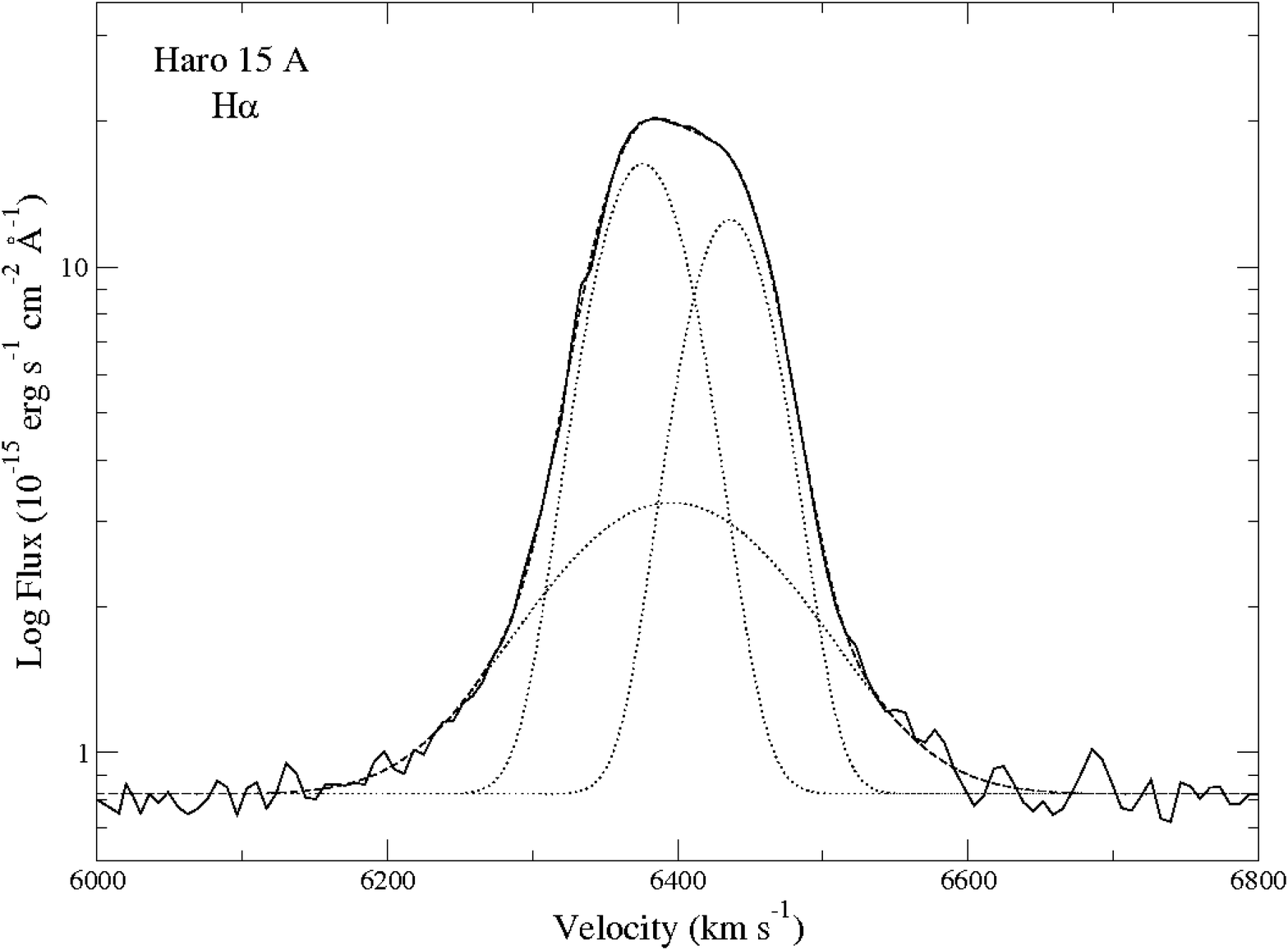}
\includegraphics[trim=0cm 0cm 0cm 0cm,clip,angle=0,width=8cm,height=5cm]{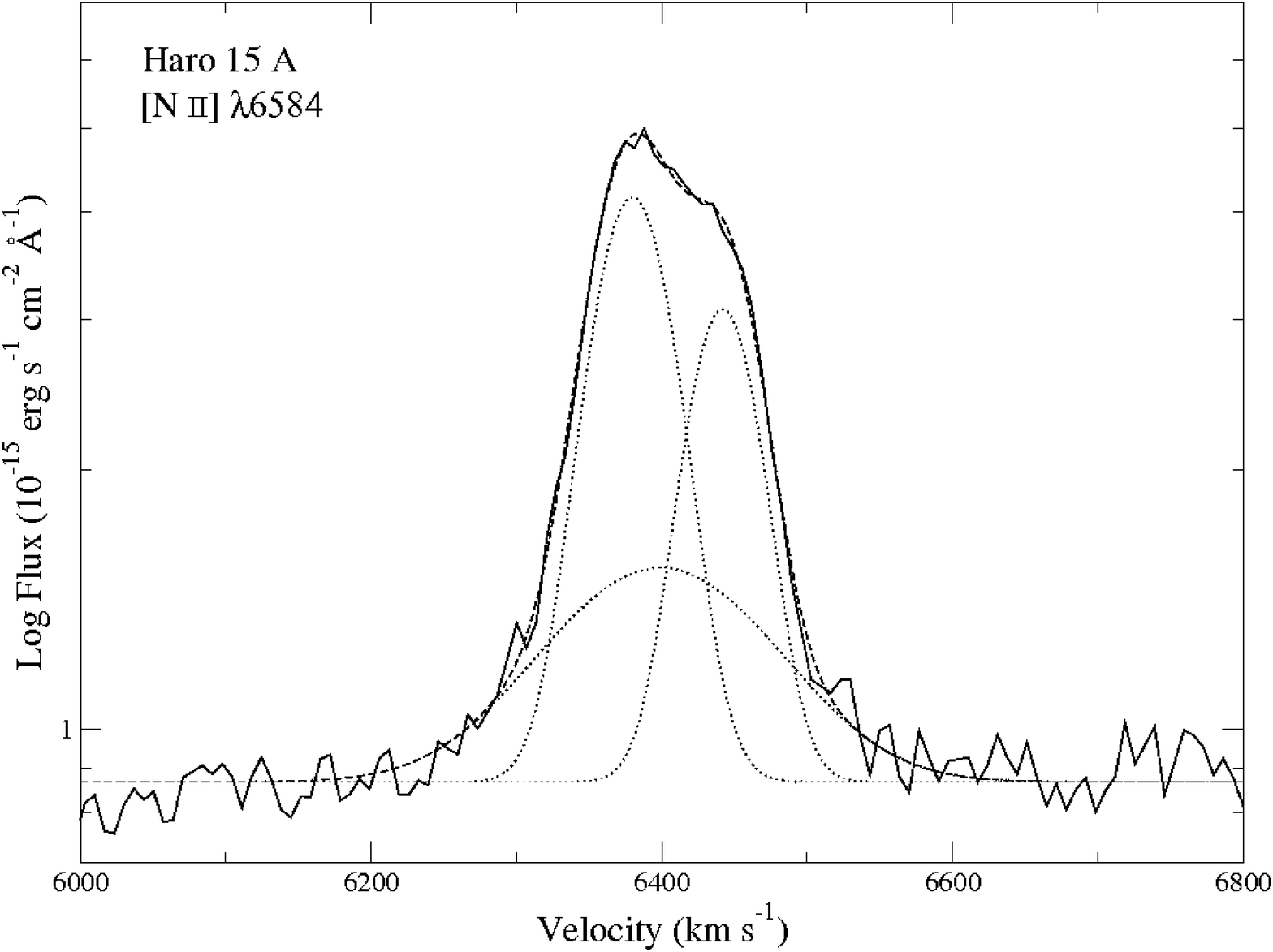}
\includegraphics[trim=0cm 0cm 0cm 0cm,clip,angle=0,width=8cm,height=5cm]{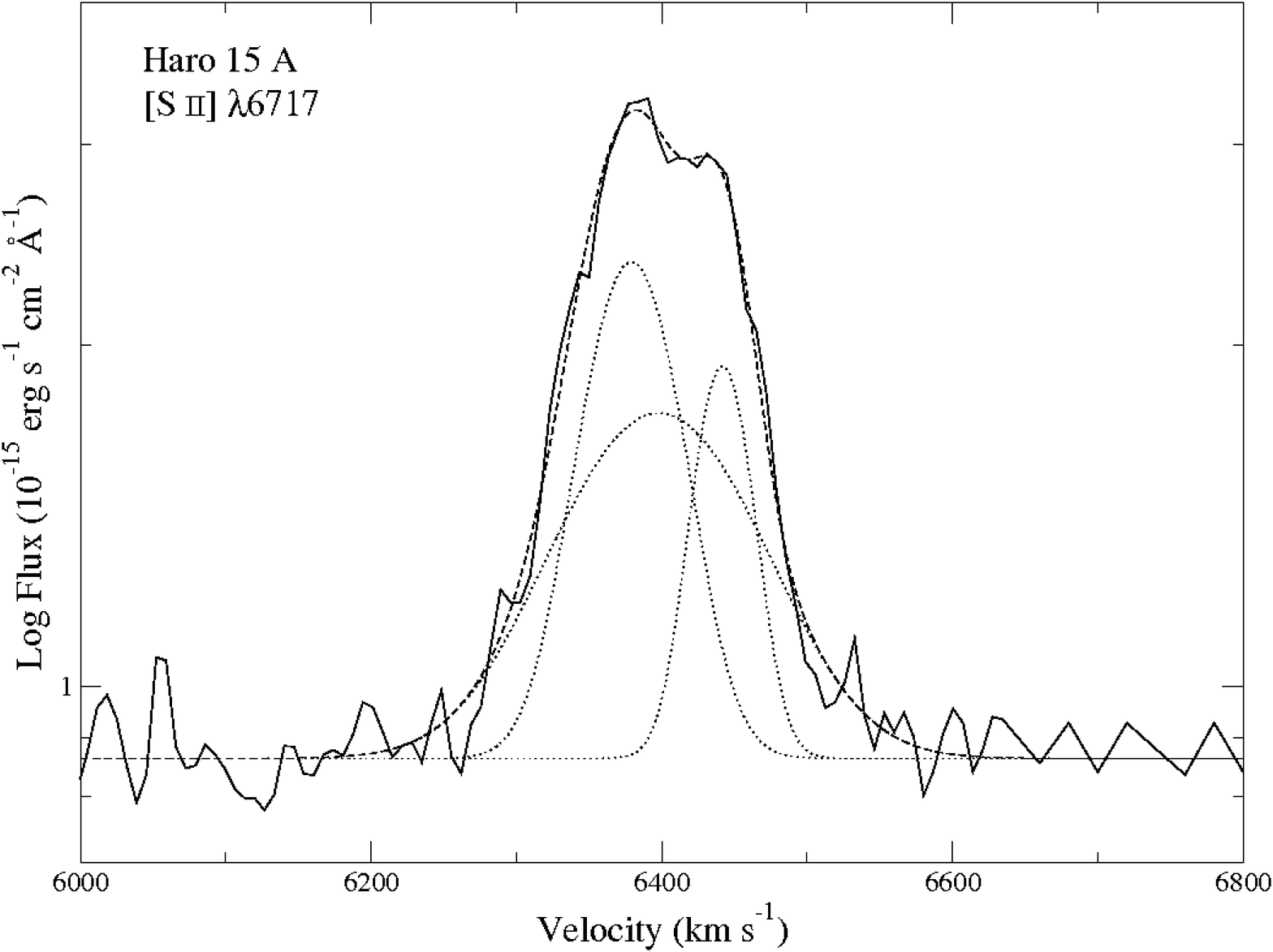}
\end{center}
\end{figure*}

\begin{figure*}
\begin{center}
\caption[VHAFig]{Same set of lines plotted in Figure \ref{figA_broad} now overlapped in the same diagram.  To enhance details at low luminosity levels, the y axes are shown in logarithmic scale. The evident similarity among different lines belonging to different elements confirms the presence of a complex kinematical structure. The results of fitting of individual components for the [\ion{S}{ii}]~$\lambda$6717 forbidden line is included as a reference. }\label{figVHA}
\includegraphics[trim=0cm 0cm 0cm 0cm,clip,angle=0,width=10cm,height=7cm]{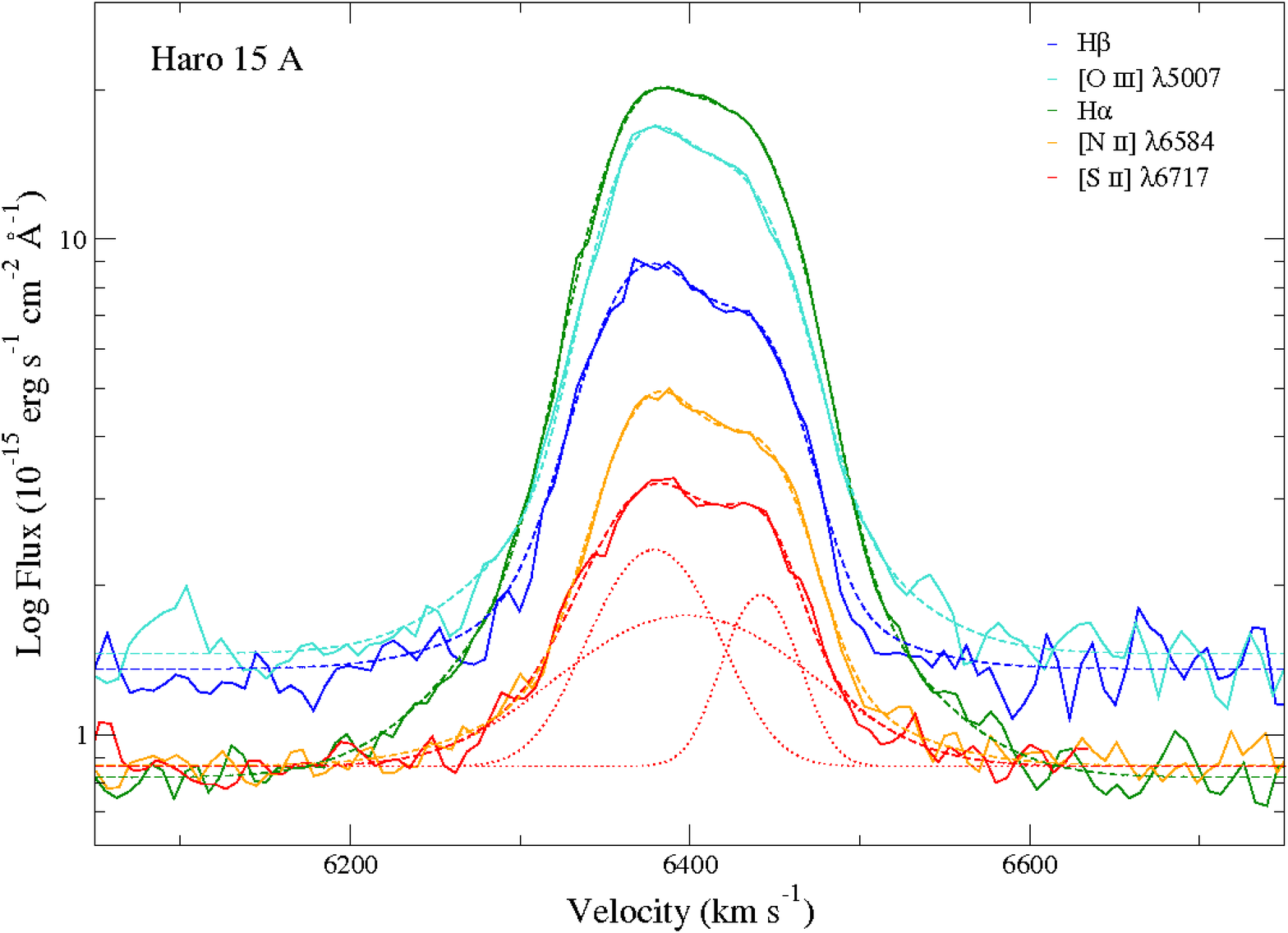}
\end{center}
\end{figure*}

We identified and fitted Gaussian profiles to the H$\beta$, [\ion{O}{iii}]$\lambda$5007, [\ion{N}{ii}]$\lambda$6548, H$\alpha$, [\ion{N}{ii}]~$\lambda$6584 and [\ion{S}{ii}]~$\lambda\lambda$6717 lines in this knot with {\sc ngaussfit} routines. The Gaussian fits of these emission line profiles are shown in Figures \ref{figA_broad} and \ref{figVHA}. The Gaussian fits in the profiles reveal the presence of two distinctly separated kinematical components labelled narrow 1 and 2 (n1 and n2). Component n1 shows a profile slightly broader than component n2, ($\sigma_{int}$ $\simeq$ 28 and 24 \kms, respectively) and both components show a relatively large spread in individual radial velocities among the different emission lines present in the spectrum. The reliability of these values is confirmed when we improve the profile fitting using {\sc ngaussfit} which in turn yields values for the profile width of each component. It is worth noting that, although the profile fitting to the [\ion{O}{iii}] lines shows similar overall results, the n2 components show the broader profile, opposite to what is found for the other emission lines. This could be related to a different kinematic behaviour of the highly ionised gas, although this needs to be confirmed for other high excitation lines.

Always considering the presence of two distinct components with different radial velocities, the overall fit continues showing the presence of a residual in the emission line wings. Following the procedures outlined in \cite{2010MNRAS.406.1094F}, we are able to fit a broad component, with a velocity dispersion of
about 78\kms\ from the H$\alpha$ emission line and slightly lower from the rest of the lines. Table \ref{tabHaro15-ABngauss} shows the parameters for the three components that fit the global profile. Individual Gaussian component fluxes are listed as fractional emission measures (EM$_{f}$) relative to the total line flux following the work by \cite{Relano05}. The sum of these individual fluxes, which we will hereafter refer to as overall H$\alpha$ flux, uncorrected for reddening, is found to be (5.49 $\pm$ 0.05) $\times$ 10$^{-14}$\ergsc. 
In Figure \ref{figA_broad} we show the {\sc ngaussfit} fitting done with three different Gaussian components in the emission lines which have enough signal to provide a reliable fit. The validity of the profile multiplicity and broadening is checked over the different emission lines, becoming more evident for the strongest emission lines. Figure \ref{figVHA} shows the excellent agreement among individual fits for the most intense emission lines in Knot A. 

\begin{figure*}
\begin{center}
\caption[AI_II-strongFig]{Top panel: Intensity distribution along the slit at the peak of the H$\alpha$ emission. Bottom panels: Result of profile extraction at the locations of regions A\,I and A\,II, together with the result of individual Gaussian profile fitting.  To enhance details at low luminosity levels, the y axes are shown in logarithmic scale. Two narrow components and a broad one are fitted for A\,I and only a single narrow and broad component are identified in A\,II.}\label{figAIAII_broad}
\includegraphics[trim=0cm 0cm 0cm 0cm,clip,angle=0,width=8cm,height=5cm]{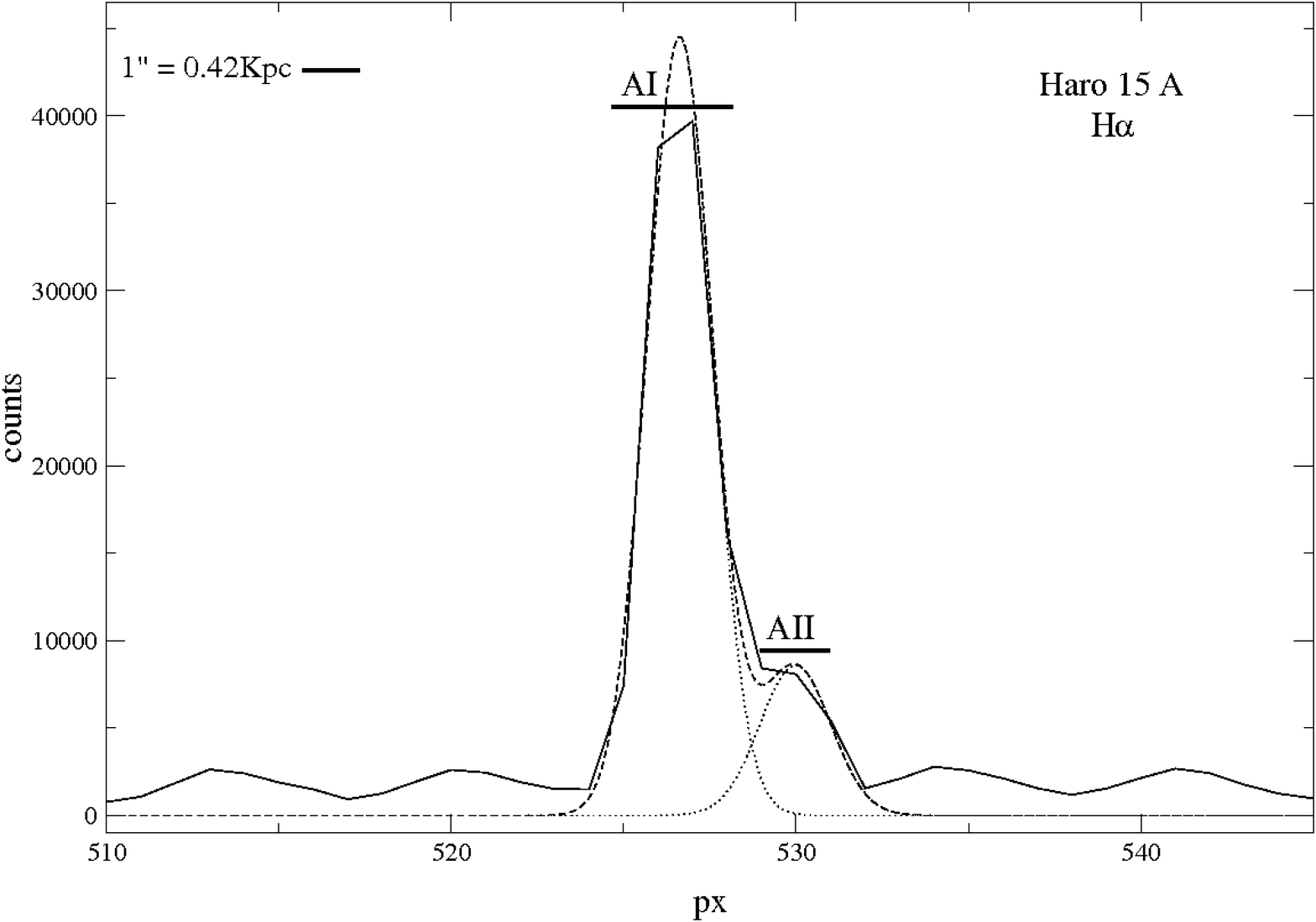}
\\
\includegraphics[trim=0cm 0cm 0cm 0cm,clip,angle=0,width=8cm,height=5cm]{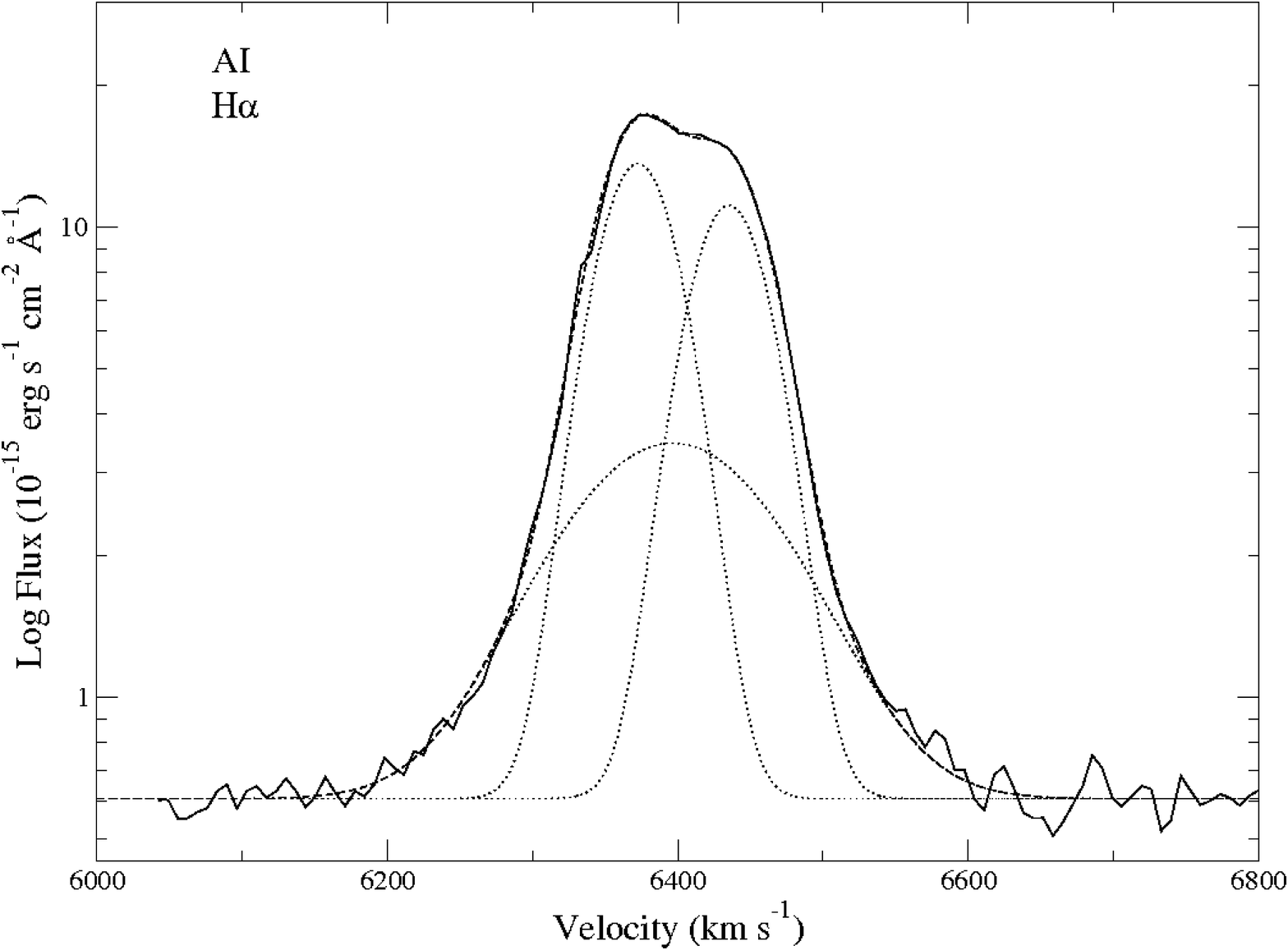}
\includegraphics[trim=0cm 0cm 0cm 0cm,clip,angle=0,width=8cm,height=5cm]{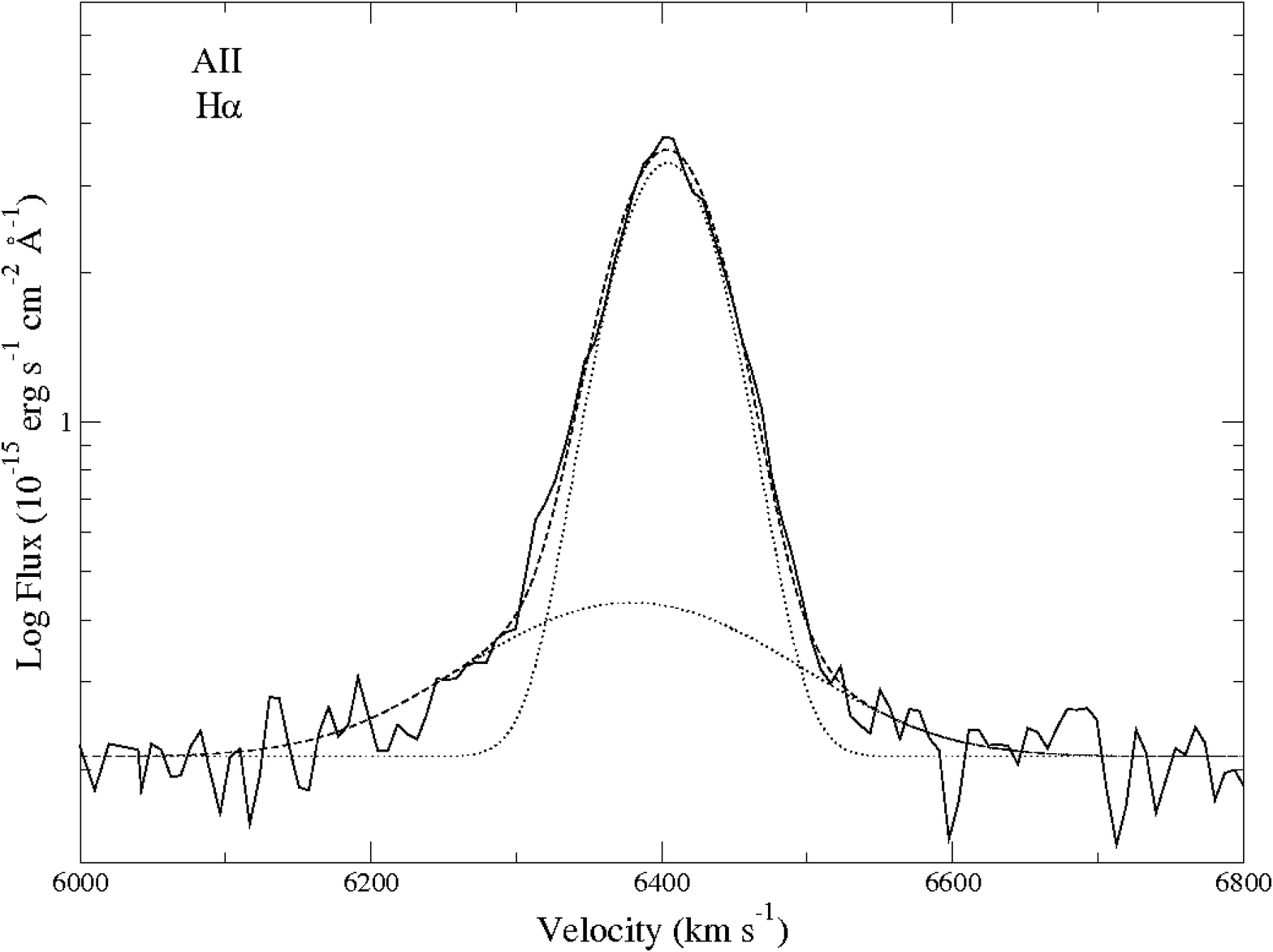}
\end{center}
\end{figure*}

The top panel in Figure \ref{figAIAII_broad} shows the spatial profile of the H$\alpha$ emission line in knot A. Two distinct zones are clearly distinguished, separated by 1.72 arcsec $\sim$ 0.72 kpc. We have performed additional extractions of these two zones (labelled A\,I and A\,II) individually. Inspection of the individual plots displayed in the bottom panels show that the kinematical structure is dominated by region A\,I, somewhat expected as it is brighter than region A\,II, which exhibits a more simple emission profile in turn. High spatial resolution integrated field spectroscopy is needed to disentangle this complex behaviour. 

\subsubsection*{\Haro\,B}

\begin{figure*}
\begin{center}
\caption[B-strongFig]{{\sc Ngaussfit} fits with two Gaussian components in
  the {\bf Haro\,15 B} strongest emission line profiles, a narrow component and a broad one. In order: H$\beta$, [\ion{O}{iii}]~$\lambda$5007, H$\alpha$, [\ion{N}{ii}]~$\lambda$6584 and [\ion{S}{ii}]~$\lambda$6717. Note that the logarithmic y-scale magnifies the actual low (1\%) error in the overall fits}\label{figB_broad}
\includegraphics[trim=0cm 0cm 0cm 0cm,clip,angle=0,width=8cm,height=5cm]{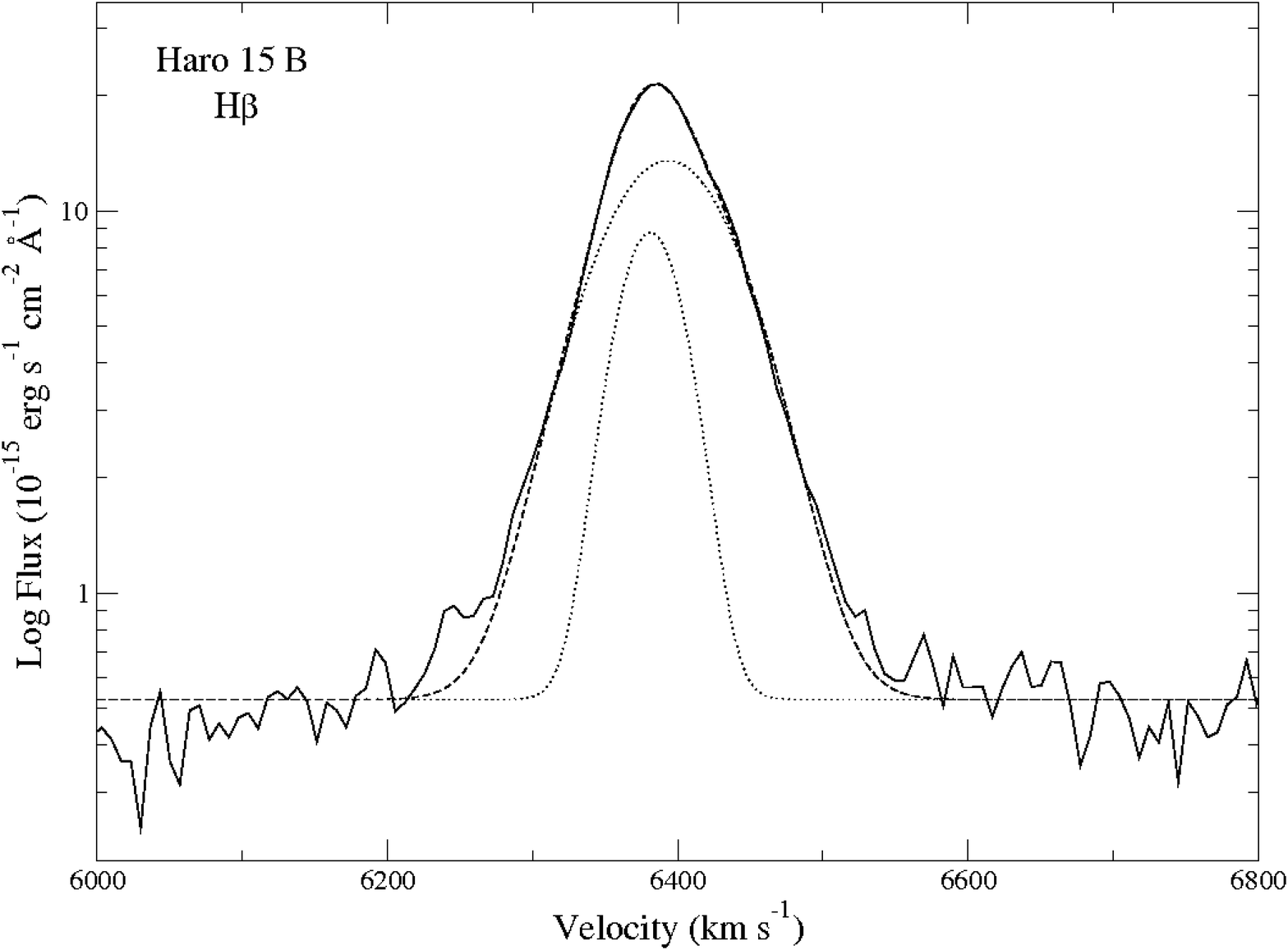}
\includegraphics[trim=0cm 0cm 0cm 0cm,clip,angle=0,width=8cm,height=5cm]{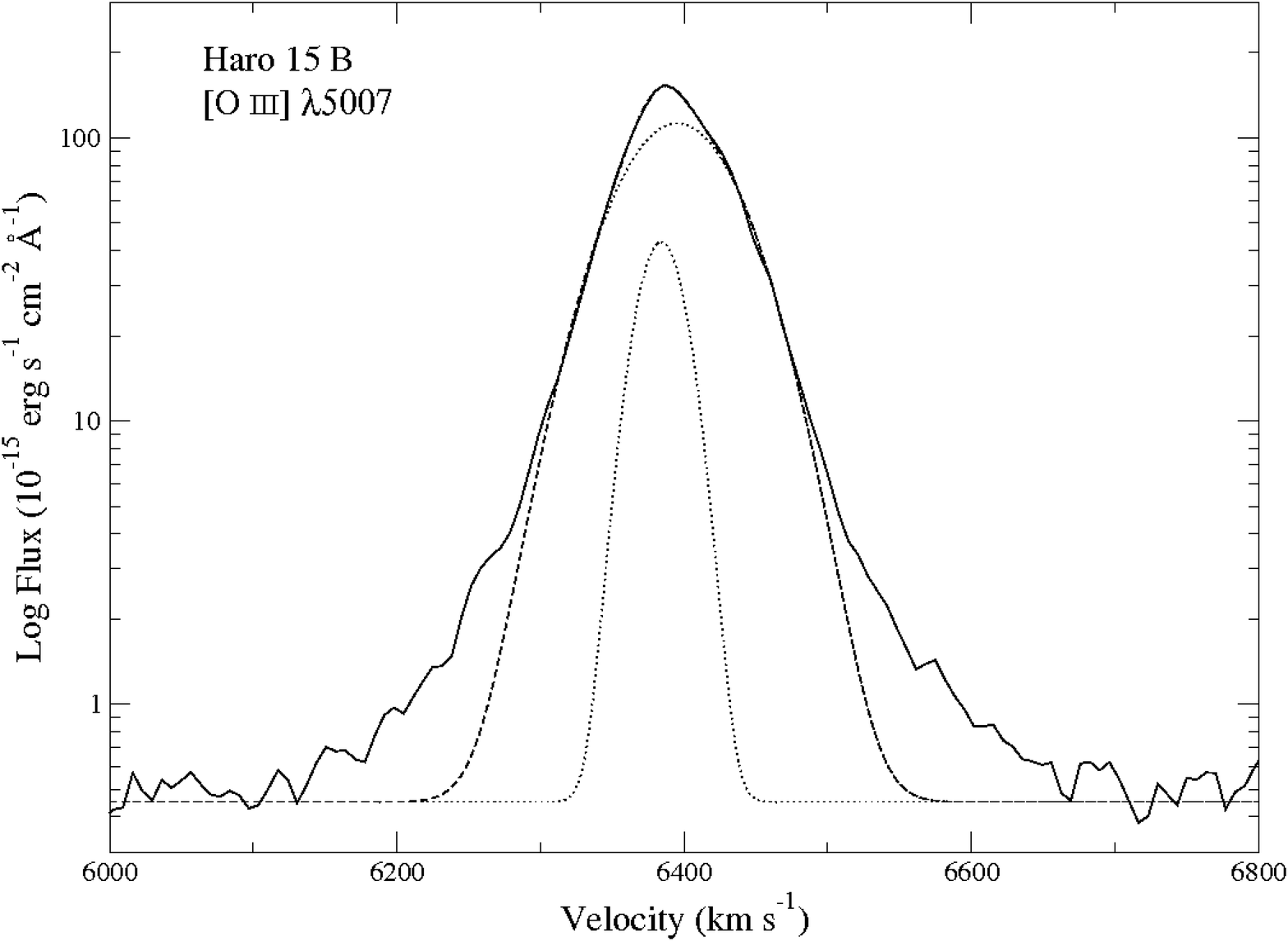}
\includegraphics[trim=0cm 0cm 0cm 0cm,clip,angle=0,width=8cm,height=5cm]{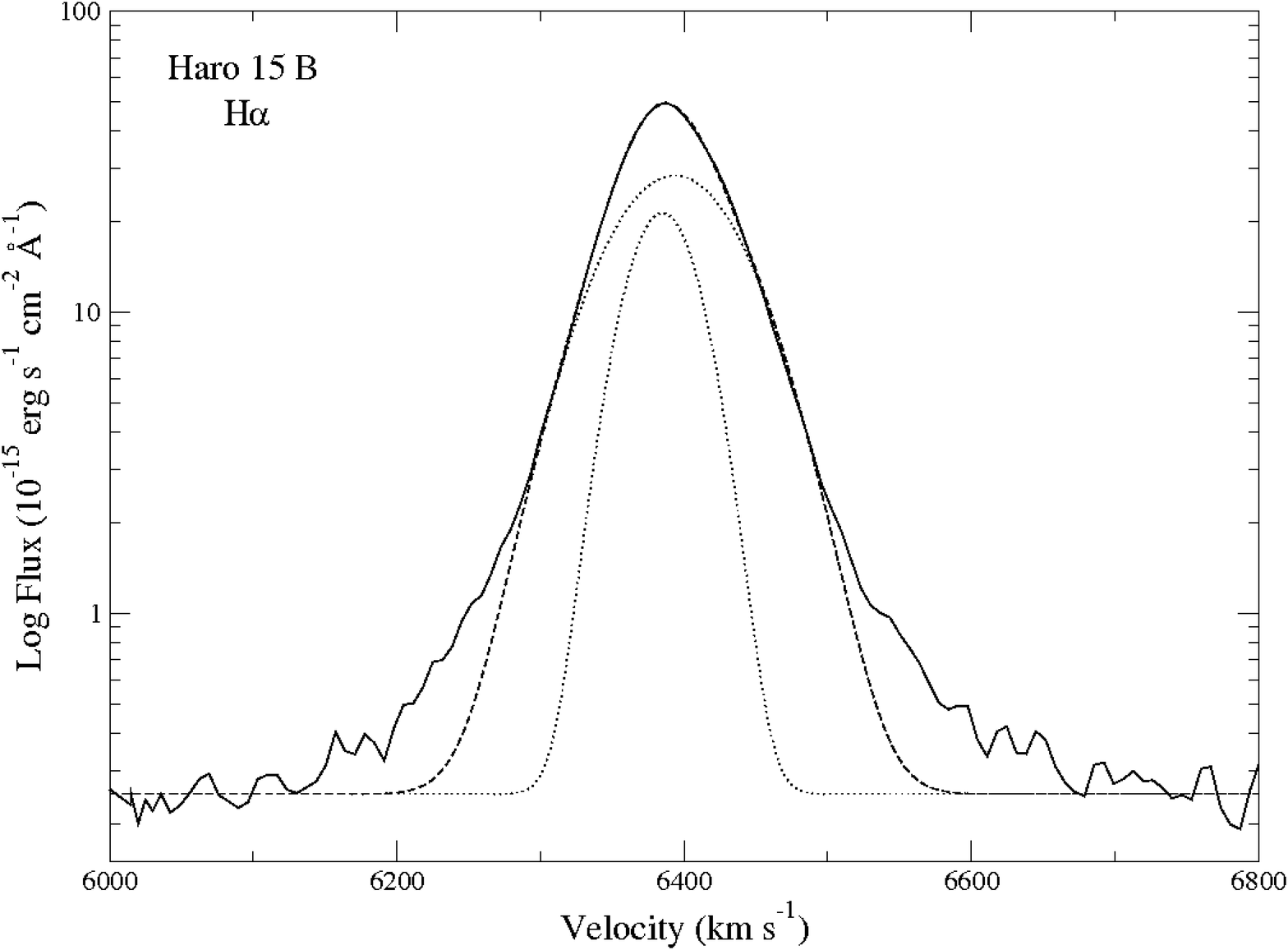}
\includegraphics[trim=0cm 0cm 0cm 0cm,clip,angle=0,width=8cm,height=5cm]{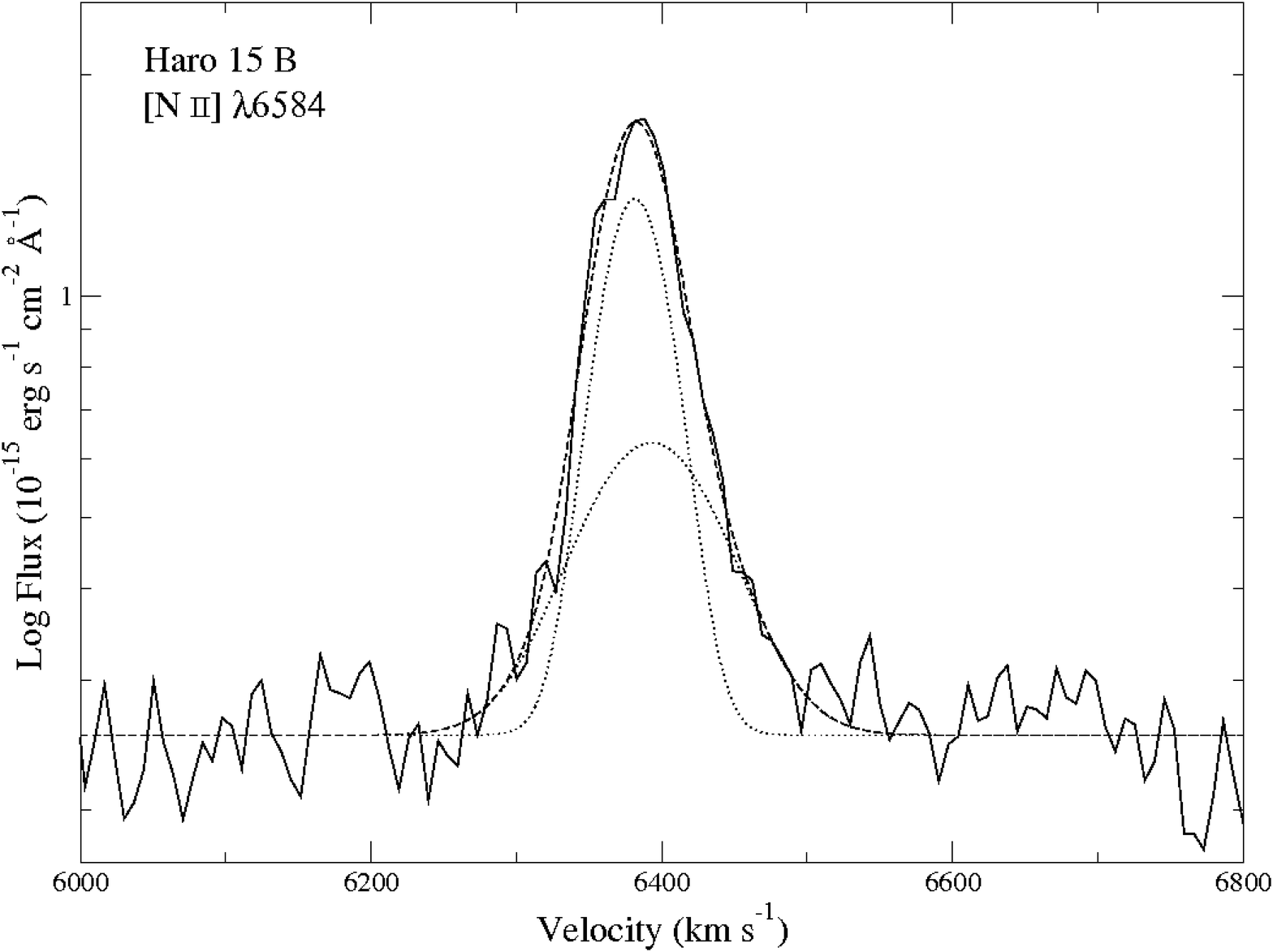}
\includegraphics[trim=0cm 0cm 0cm 0cm,clip,angle=0,width=8cm,height=5cm]{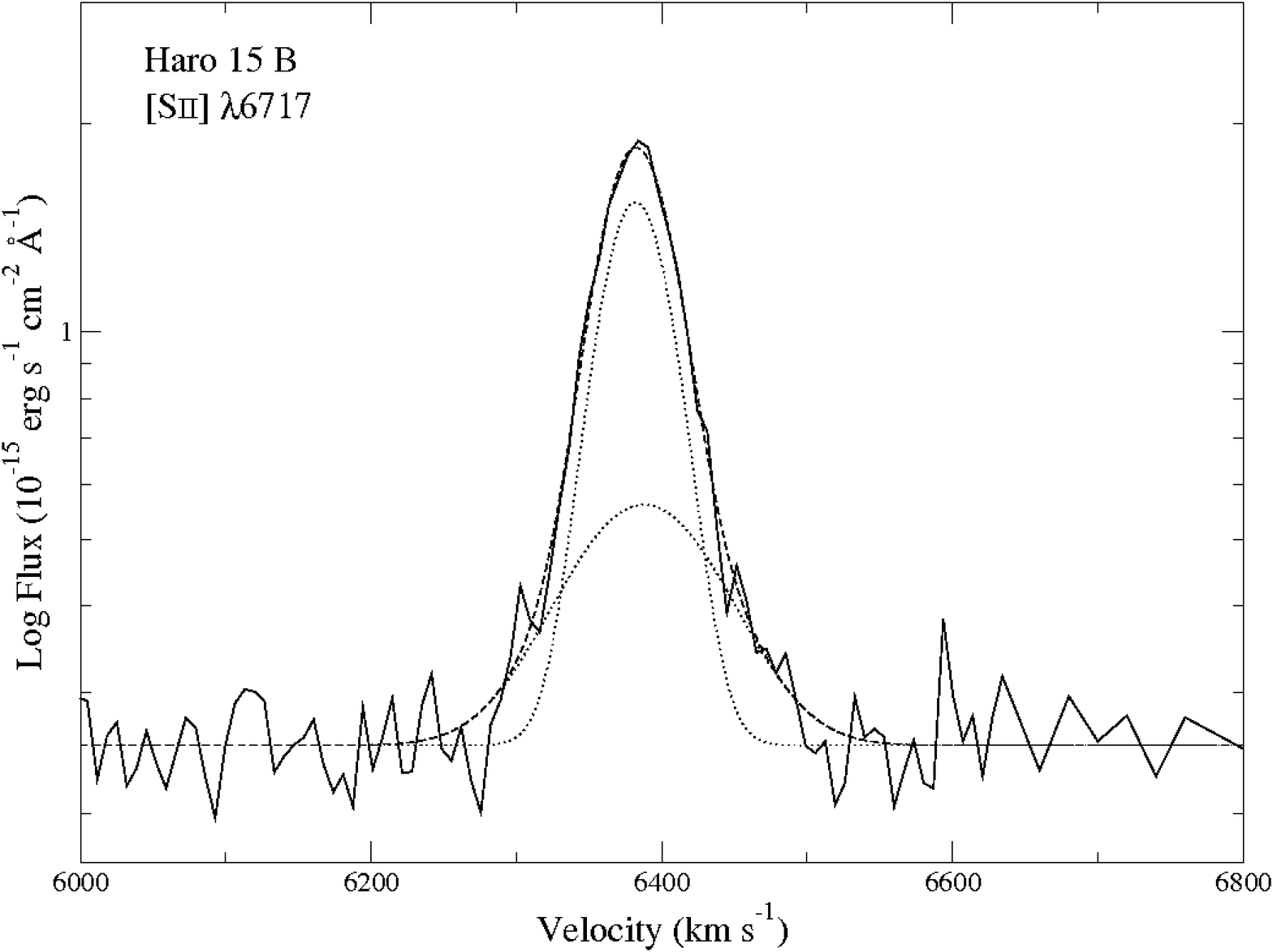}
\end{center}
\end{figure*}

The spectrum of region B is dominated by nebular emission lines, where we detected and measured [\ion{O}{iii}]~$\lambda$4363 and \ion{He}{ii}~$\lambda$4686. We identified and fitted Gaussian profiles to the H$\beta$, [\ion{O}{iii}]~$\lambda\lambda$4959,5007, [\ion{N}{ii}]~$\lambda$6548, H$\alpha$,
[\ion{N}{ii}]~$\lambda$6584 and [\ion{S}{ii}]~$\lambda\lambda$6717,6731 lines in this region using {\sc ngaussfit}.
It is interesting to note that, although from inspection of the Wide Field Planetary Camera 2 image shown in Figure \ref{figHaro15chart}, a complex structure of region B  is apparent, this is not reflected in the kinematical behaviour of the region, since it is possible to fit a single Gaussian component with a velocity dispersion of 20 \kms. Nevertheless, the profile fittings show the presence of a residual in the emission line wings that we were able to explain by fitting a broad component with a velocity dispersion of about 43 \kms\ Results of the fitting procedure can be seen in Figure \ref{figB_broad}; it is worth noting that the logarithmic scale chosen for the y-axis magnifies the 1\% residual in the profile wings for  H$\alpha$ and [\ion{O}{iii}]~$\lambda$5007. 
For the listed lines the derived radial velocities, velocity dispersions and their corresponding errors are listed in Table \ref{tabHaro15-ABngauss}. The overall H$\alpha$ flux, uncorrected for reddening, is found to be (9.88 $\pm$ 0.3) $\times$ 10$^{-14}$\ergsc.

As concluded by \cite{2009A&A...508..615L}, knot B has a much higher ionization degree than the rest of the regions in \Haro. These authors suggest that this may be a consequence of the extreme youth of this knot. This fact and the low metallicity found in knot B (lower than for the other regions) indicate that it has a different nature, probably being the remnant of a dwarf galaxy which is experiencing a minor interaction with \Haro\ \citep{2010A&A...521A..63L}.

\begin{table*}
\caption[Haro15-ABcomp]{Results of Gaussian profiles fitting to the observed emission lines in {\bf \Haro\, A} and {\bf B}. Each emission line is identified by its ion laboratory wavelength and ion name in columns 1 and 2. According to the different fits performed on each line, column 3 identifies each ``narrow component (1 and 2, where applicable), and a broad component. Radial velocities (V$_r$) and intrinsic velocity dispersions ($\sigma_{int}$) together with their respective errors are expressed in \kms. The intrinsic velocity dispersions are corrected for the instrumental and thermal widths. Fractional emission measures (EM$_{f}$) in $\%$.}
\label{tabHaro15-ABngauss}
\begin{center}
\begin{tabular}{lllllllllllll}
\hline
\multicolumn{3}{c}{} & \multicolumn{5}{c}{\em \Haro\, A} &  \multicolumn{5}{c}{\em \Haro\, B} \\
$\lambda_{0}$&ion&comp.&V$_r$&error&$\sigma_{int}$&error&EM$_{f}$&V$_r$&error&$\sigma_{int}$&error&EM$_{f}$\\
\hline
4861	&	H$\beta$	&	narrow 1	&	6375.5	&	0.8	&	28.8	&	1.1	&	55.7	&	6381.2	&	0.8	&	17.3	&	1.1	&	23.6	\\
	&		&	narrow 2	&	6440.1	&	0.9	&	20.7	&	1.1	&	25.6	&	\ldots	&	\ldots	&	\ldots	&	\ldots	&	\ldots	\\
	&		&	broad	&	6390.6	&	5.2	&	72.5	&	2.4	&	18.7	&	6393.4	&	0.8	&	43.3	&	1.1	&	76.4	\\
\\																									
4959	&	[\ion{O}{iii}]	&	narrow 1	&	6373.4	&	1.1	&	26.5	&	1.3	&	39.0	&	6383.2	&	0.8	&	16.5	&	1.1	&	15.5	\\
	&		&	narrow 2	&	6426.2	&	1.3	&	31.9	&	1.5	&	38.0	&	\ldots	&	\ldots	&	\ldots	&	\ldots	&	\ldots	\\
	&		&	broad	&	6384.4	&	4.7	&	64.0	&	6.0	&	23.0	&	6394.4	&	0.7	&	40.1	&	1.1	&	84.5	\\
\\																									
5007	&	[\ion{O}{iii}]	&	narrow 1	&	6372.7	&	0.7	&	24.9	&	1.1	&	39.2	&	6384.1	&	0.8	&	15.1	&	1.1	&	13.0	\\
	&		&	narrow 2	&	6429.1	&	0.8	&	28.8	&	1.1	&	35.4	&	\ldots	&	\ldots	&	\ldots	&	\ldots	&	\ldots	\\
	&		&	broad	&	6396.2	&	1.6	&	73.0	&	2.4	&	25.4	&	6395.2	&	0.7	&	40.2	&	1.0	&	87.0	\\
\\																									
6548	&	[\ion{N}{ii}]	&	narrow 1	&	6378.7	&	1.2	&	28.6	&	1.1	&	53.4	&	6381.9	&	1.3	&	20.6	&	1.2	&	50.2	\\
	&		&	narrow 2	&	6438.3	&	1.1	&	21.5	&	1.3	&	27.9	&	\ldots	&	\ldots	&	\ldots	&	\ldots	&	\ldots	\\
	&		&	broad	&	6405.5	&	12.1	&	72.0	&	14.1	&	18.7	&	6395.6	&	4.2	&	49.0	&	4.7	&	49.8	\\
\\																									
6563	&	H$\alpha$	&	narrow 1	&	6376.1	&	0.6	&	27.3	&	0.8	&	47.6	&	6385.0	&	0.6	&	20.0	&	0.8	&	28.4	\\
	&		&	narrow 2	&	6436.0	&	0.6	&	24.5	&	0.8	&	33.0	&	\ldots	&	\ldots	&	\ldots	&	\ldots	&	\ldots	\\
	&		&	broad	&	6395.8	&	2.6	&	77.8	&	1.7	&	19.4	&	6393.5	&	0.6	&	43.7	&	0.8	&	71.6	\\
\\																									
6584	&	[\ion{N}{ii}]	&	narrow 1	&	6379.9	&	0.6	&	27.5	&	0.8	&	46.6	&	6381.5	&	0.7	&	24.3	&	0.9	&	59.8	\\
	&		&	narrow 2	&	6441.9	&	0.7	&	24.5	&	0.9	&	28.1	&	\ldots	&	\ldots	&	\ldots	&	\ldots	&	\ldots	\\
	&		&	broad	&	6399.2	&	2.1	&	74.7	&	2.5	&	25.3	&	6392.9	&	2.3	&	48.4	&	2.3	&	40.2	\\
\\																									
6717	&	[\ion{S}{ii}]	&	narrow 1	&	6379.1	&	1.2	&	31.7	&	0.9	&	38.1	&	6382.0	&	0.8	&	25.1	&	0.8	&	67.1	\\
	&		&	narrow 2	&	6441.8	&	1.2	&	19.1	&	1.3	&	16.5	&	\ldots	&	\ldots	&	\ldots	&	\ldots	&	\ldots	\\
	&		&	broad	&	6397.9	&	2.2	&	65.5	&	2.4	&	45.5	&	6387.8	&	3.2	&	52.4	&	3.3	&	32.9	\\
\\																									
6731	&	[\ion{S}{ii}]	&	narrow 1	&	6378.3	&	1.5	&	32.3	&	1.7	&	37.2	&	6384.9	&	0.8	&	23.4	&	1.0	&	66.6	\\
	&		&	narrow 2	&	6442.3	&	1.1	&	16.7	&	1.3	&	20.2	&	\ldots	&	\ldots	&	\ldots	&	\ldots	&	\ldots	\\
	&		&	broad	&	6398.3	&	3.4	&	65.3	&	4.3	&	42.6	&	6380.6	&	4.8	&	59.0	&	5.3	&	33.4	\\
\hline 
\end{tabular}
\end{center}
\end{table*}

\subsubsection*{\Haro\,C}

Although the observed spectrum of region C is faint, we were able to identify and fit Gaussian profiles to the H$\beta$, [\ion{O}{iii}]~$\lambda$5007, H$\alpha$, [\ion{N}{ii}]~$\lambda$6584 and [\ion{S}{ii}]~$\lambda\lambda$6717,6731 lines. 
Knot C shows one narrow component together with the underlying broad component. However, due to the poor signal of the spectrum, all emission lines were fitted using the H$\alpha$ {\sc ngaussfit} solution as a template, fixing the profile centers and widths, and only allowing the task to fit the profile amplitudes. 
In Figure \ref{figCEF_broad} we only show H$\alpha$ since the rest of the intense lines are very noisy due to the low signal-to-noise ratio of the spectrum. Table \ref{tabHaro15-CEFngauss} shows the derived kinematical parameters. The overall H$\alpha$ flux, uncorrected for reddening is (2.67 $\pm$ 0.07) $\times$ 10$^{-15}$ \ergsc.


\subsubsection*{\Haro\,E}
We detect the presence of a broad profile, suggesting the presence of a double peak together with a flux excess in the strongest emission lines. We identified and fitted Gaussian profiles to the H$\beta$, [\ion{O}{iii}]~$\lambda$5007, H$\alpha$, [\ion{N}{ii}]~$\lambda$6584 and [\ion{S}{ii}]~$\lambda\lambda$6717,6731 lines. The results of the profile fitting procedure yields two components of similar width. Although both are supersonic profiles it is not possible to distinguish a broad and narrow component and therefore refer to the identified components as narrow 1 and 2 (see Figure \ref{figCEF_broad}). For the listed lines the derived radial velocities, the velocity dispersions and the corresponding errors are listed in Table \ref{tabHaro15-CEFngauss}. The overall H$\alpha$ flux, uncorrected for reddening, is found to be (4.99 $\pm$ 0.14) $\times$ 10$^{-15}$\ergsc.

Owing to the poor signal-to-noise spectrum we fitted some of the  emission lines such as [\ion{N}{ii}]~$\lambda$6584 and [\ion{S}{ii}]~$\lambda\lambda$6717,6731, using the H$\alpha$ {\sc ngaussfit} solution, and iterating separately for individual set of parameters, following the procedure described in \cite{2010MNRAS.406.1094F} .


\subsubsection*{\Haro\,F}
\begin{figure*}
\begin{center}
\caption[CEF-strongFig]{{\sc Ngaussfit} fits with three Gaussian components in
  the {\bf Haro\,15 C, E,} and {\bf F} in H$\alpha$ emission line profiles, a narrow component and a broad one.}\label{figCEF_broad}
\includegraphics[trim=0cm 0cm 0cm 0cm,clip,angle=0,width=8cm,height=5cm]{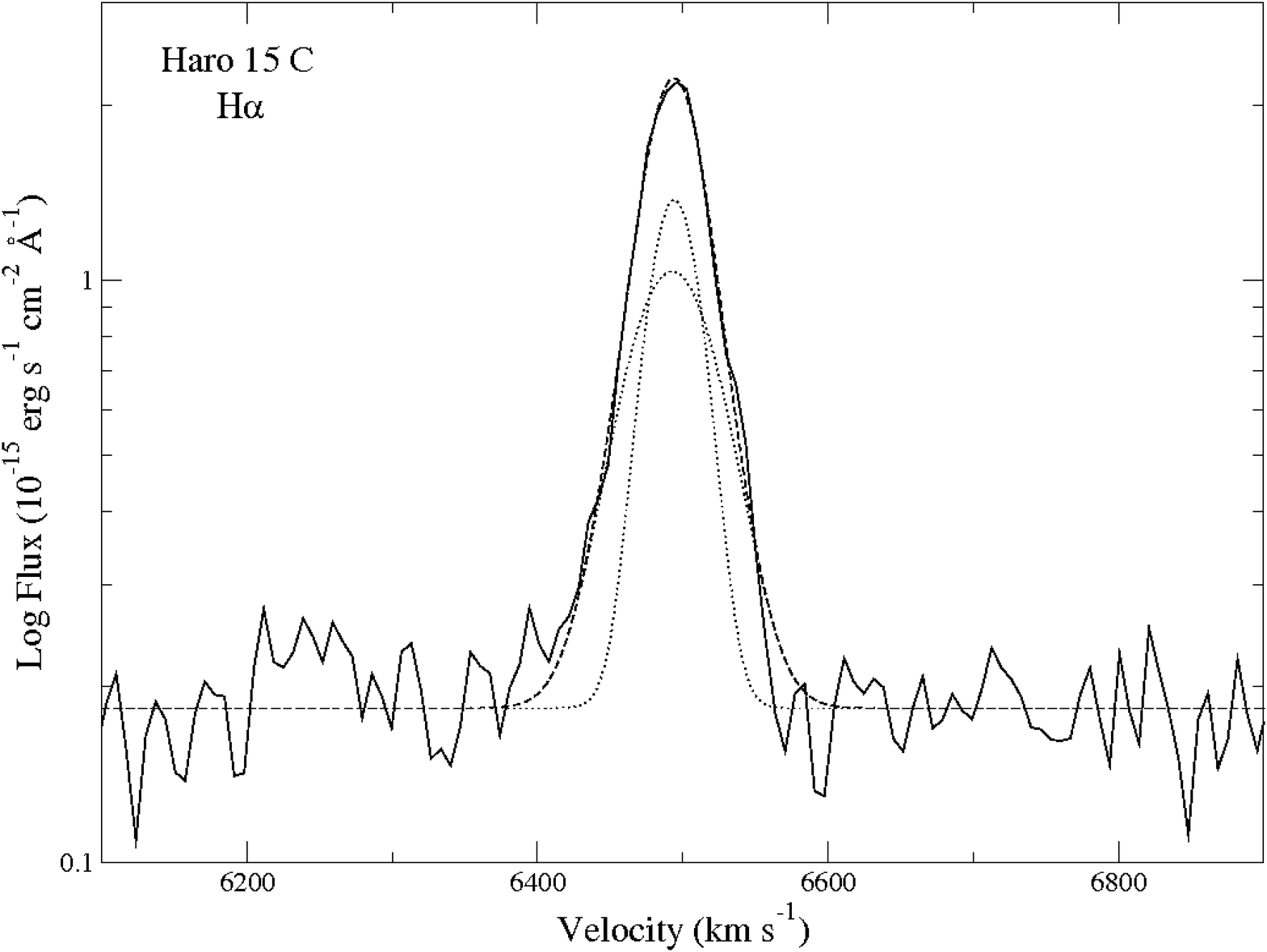}
\includegraphics[trim=0cm 0cm 0cm 0cm,clip,angle=0,width=8cm,height=5cm]{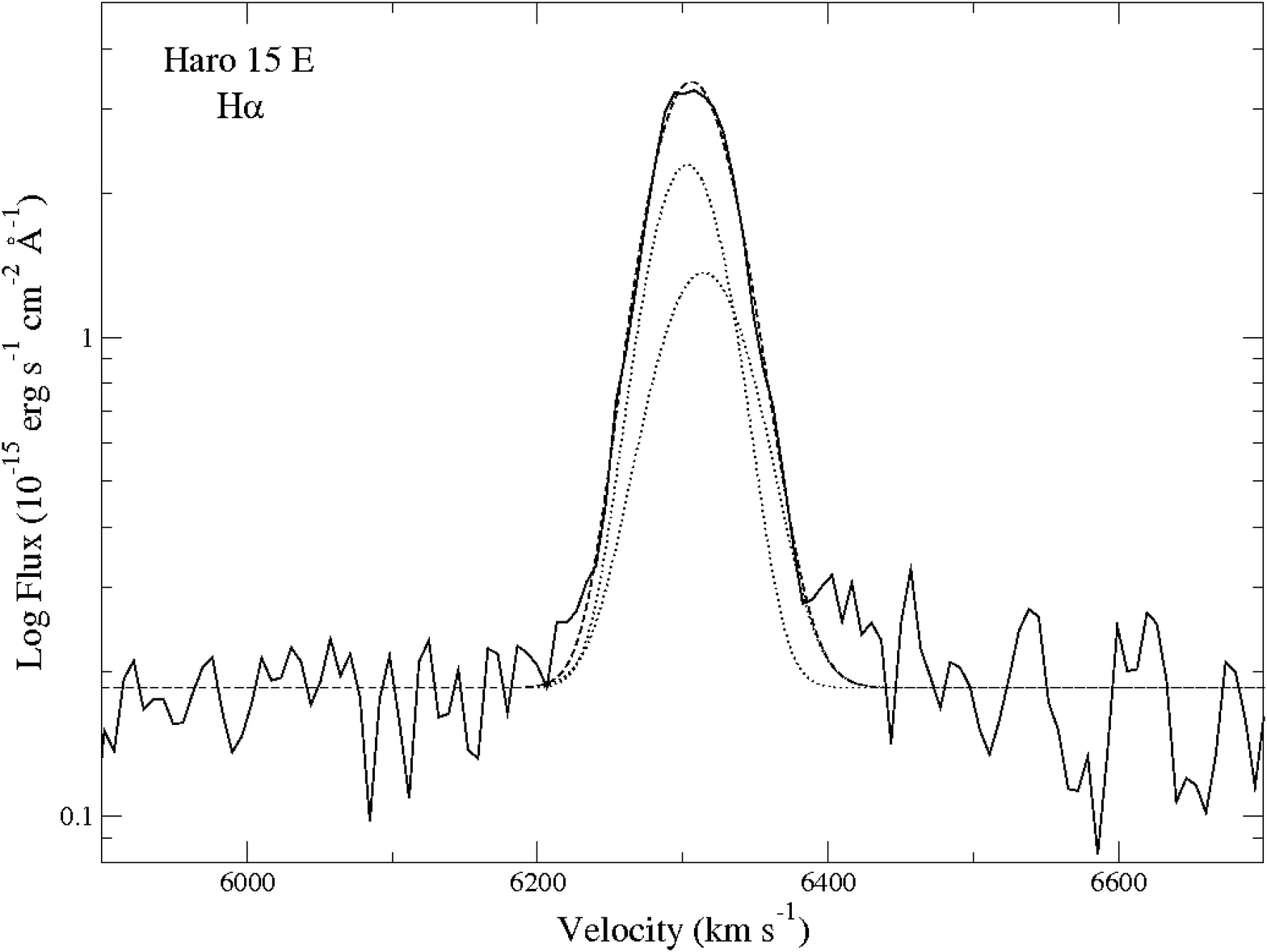}
\includegraphics[trim=0cm 0cm 0cm 0cm,clip,angle=0,width=8cm,height=5cm]{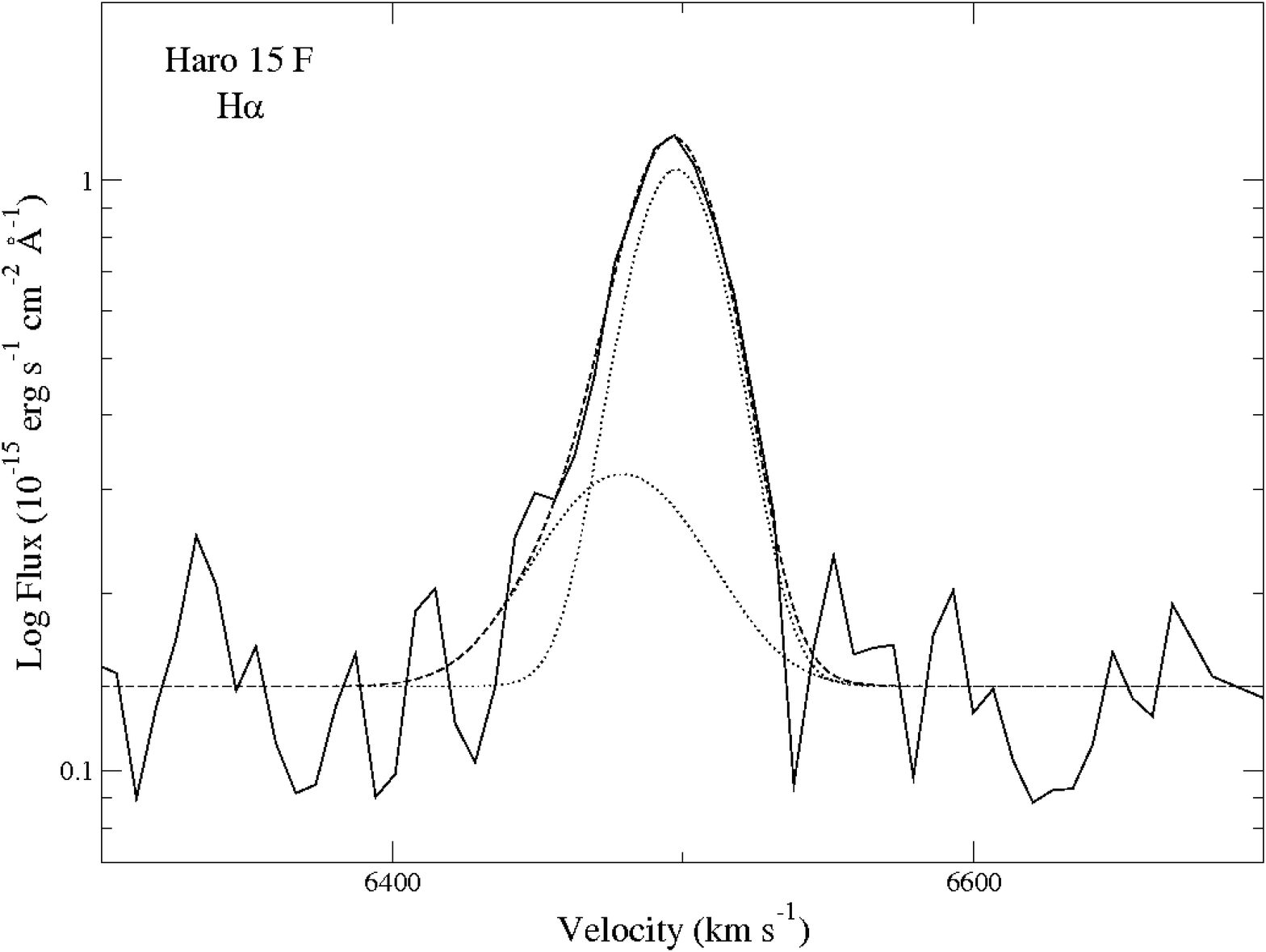}
\end{center}
\end{figure*}

The observed spectrum of knot F is the weakest one, and hence we only identified and fitted the Gaussian profiles of the H$\beta$, [\ion{O}{iii}]~$\lambda$5007, and H$\alpha$ lines. H$\beta$ and [\ion{O}{iii}]~$\lambda$5007 are in fact very noisy and narrow, rendering  impossible the task of fitting multiple Gaussian components (narrow plus broad). 
The spectral line broadening in their integrated spectra is only observed in the H$\alpha$ line where we could fit a narrow Gaussian component together with the broad one, with velocity dispersions of about 8.3 \kms\ and 22 \kms, respectively (see Figure \ref{figCEF_broad}). The estimated velocity dispersion for the narrow component in the H$\alpha$ line is subsonic (8 \kms), typical of classic \HII\ regions, although the presence of a broad component is almost exclusive of Giant H\,{\sc ii} Regions.

For H$\alpha$, the derived radial velocity, the velocity dispersions and the corresponding errors are listed in Table \ref{tabHaro15-CEFngauss}. The overall H$\alpha$ flux, uncorrected for reddening, is found to be (1.06 $\pm$ 0.16) $\times$ 10$^{-15}$\ergsc.

\begin{table*}
\caption[Haro15-CEFcomp]{Results of Gaussian profiles fitting to the observed emission lines in {\bf \Haro\, C, E,} and {\bf F}. Each emission line is identified by its ion laboratory wavelength and ion name in columns 1 and 2. According to the different fits performed on each line, column 3 identifies the narrow and  broad components. Radial velocities (V$_r$) and intrinsic velocity dispersions ($\sigma_{int}$) together with their respective errors are expressed in \kms. The intrinsic velocity dispersions are corrected for the instrumental and thermal widths. Fractional emission measures (EM$_{f}$)  in $\%$.}
\label{tabHaro15-CEFngauss}
\begin{center}
\begin{tabular}{llllllll}
\hline
$\lambda_{0}$&ion&comp.&V$_r$&error&$\sigma_{int}$&error&EM$_{f}$\\
\hline
\multicolumn{3}{c}{} & \multicolumn{5}{l}{\em \Haro\, C}  \\
\\
6563	&	H$\alpha$	&	narrow 	&	6494.2	&	0.4	&	11.8	&	0.9	&	44.5	\\
	&		&	broad	&	6492.7	&	0.7	&	28.5  &1.0  &  55.5 \\
\\
\multicolumn{3}{c}{} & \multicolumn{5}{l}{\em \Haro\, E}  \\
\\
4861	&	H$\beta$	&	narrow 1	&	6303.0	&	0.8	&	21.4	&	1.2	&	48.6	\\
	&		&	narrow 2	&	6314.4	&	1.1	&	28.2	&	1.4	&	51.4	\\
\\															
5007	&	[\ion{O}{iii}]	&	narrow  1	&	6302.9	&	1.7	&	23.9	&	2.2	&	32.8	\\
	&		&	narrow 2	&	6307.6	&	1.2	&	27.6	&	1.5	&	67.2	\\
\\															
6563	&	H$\alpha$	&	narrow  1	&	6303.0	&	0.7	&	21.4	&	0.9	&	59.1	\\
	&		&	narrow 2	&	6314.4	&	1.0	&	28.2	&	1.2	&	40.9	\\
\\															
6584	&	[\ion{N}{ii}]	&	narrow 1 	&	6306.3	&	2.0	&	22.4	&	2.9	&	31.2	\\
	&		&	narrow 2	&	6312.4	&	2.0	&	34.2	&	2.0	&	68.8	\\
\\															
6717	&	[\ion{S}{ii}]	&	narrow 1	&	6309.0	&	1.9	&	17.6	&	1.6	&	24.9	\\
	&		&	narrow 2	&	6314.3	&	1.4	&	27.8	&	1.5	&	75.1	\\
\\															
6731	&	[\ion{S}{ii}]	&	narrow 1 	&	6306.1	&	3.5	&	14.9	&	2.3	&	29.6	\\
	&		&	narrow 2	&	6314.7	&	3.4	&	30.0	&	2.9	&	70.4	\\
\\
\multicolumn{3}{c}{} & \multicolumn{5}{l}{\em \Haro\, F} \\
\\
6563	&	H$\alpha$	&	narrow 	&		6497.7	&	1.1	&	8.3	&	1.0	&	75.9	\\
	&		&	broad	&		6479.1	&	3.4	&	22.0	&	5.7	&	24.1	\\
\hline
\end{tabular}
\end{center}
\end{table*}

\subsection{Radial Velocities}
\label{sec:Radial Velocities} 

\begin{figure*}
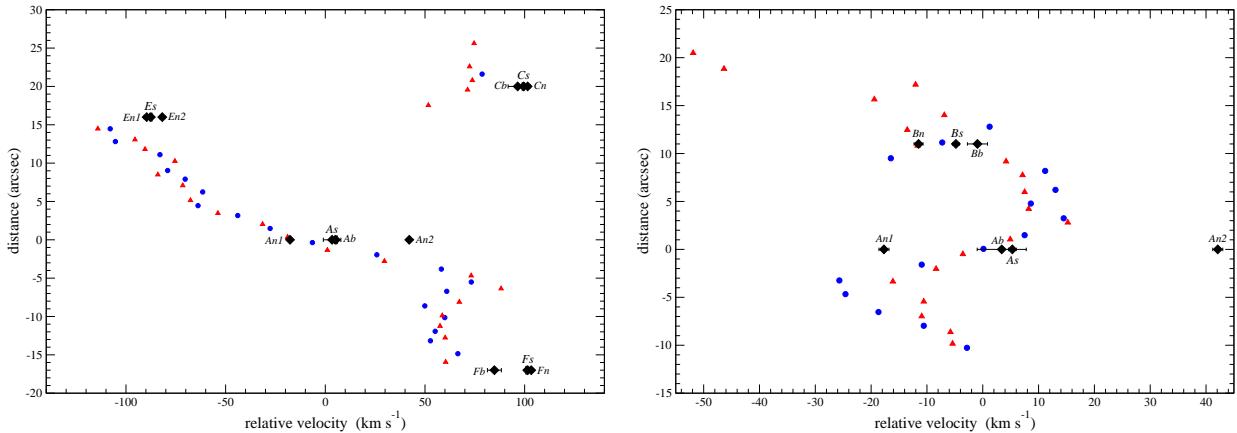

\begin{center}
\caption[Pos-vel41-117]{Position vs.\ Velocity diagrams for the observed knots in \Haro. H$\alpha$  radial velocity results, relative to knot A, are over imposed on the data by \cite{2009A&A...508..615L} for a slit positions PA 41$^{\circ }$ (left panel) and  PA 117$^{\circ }$(right panel). The H$\alpha$ and [\ion{O}{iii}]~5007 data of \cite{2009A&A...508..615L} in \Haro\ are identified by blue circles and red triangles, respectively.}
\label{figPos-vel41-117}
\includegraphics[angle=0,width=.45\textwidth]{P41_vel_comp.eps} \hspace{0.2cm}
\includegraphics[angle=0,width=.45\textwidth]{P117_vel_comp.eps}
\end{center}
\end{figure*}

The radial velocities for each component of the different knots of \Haro\ are given in the corresponding tables where the results for the profile fits are shown. In Table \ref{tabVr} we list the average radial velocity for each component, together with the corresponding errors. The single component is the result of a single Gaussian fit to the emission line profiles.
In order to compare our results with those of \cite{2009A&A...508..615L} we have plotted our radial velocities in Figure \ref{figPos-vel41-117} relative to the velocity of the galactic center 6415 \kms\ \citep{2006PhDT........35L}, after correcting to the Galactic Standard of Rest (GSR) using the {\sc rvcorrect} task by {\sc IRAF}.
Figure \ref{figPos-vel41-117} shows our results together with data presented by \cite{2009A&A...508..615L} for their long-slits with position angles (PA) of 41$^{\circ }$ and 117$^{\circ }$. Regarding the single component velocity, it can be clearly seen that the average radial velocities derived from the emission lines follow nicely those derived by \cite{2009A&A...508..615L}. However, it is worth mentioning that individual velocities of narrow components (n1 and n2) for knot A differ by about 60 \kms. This cannot be easily explained by scatter within galactic rotation, suggesting that their relative velocities are due to mutual orbital motion around their gravity center.

\begin{table*}
\caption[Observed velocities]{Average LSR radial velocities for \Haro\ emission line knots. Column 1 indicates the fitted feature and the remaining columns list the average radial velocity for each knot and its uncertainty, both in \kms. }
\label{tabVr}
\begin{center}
\begin{tabular}{@{}llccccccccc@{}}
\hline
Comp.& $<Vr>$ & error & $<Vr>$ & error & $<Vr>$ & error& $<Vr>$ & error & $<Vr>$ & error\\
\hline
 &\multicolumn{2}{c}{\em \Haro\,A} & \multicolumn{2}{c}{\em \Haro\,B}&  \multicolumn{2}{c}{\em \Haro\,C}&\multicolumn{2}{c}{\em \Haro\,E} & \multicolumn{2}{c}{\em \Haro\,F}\\   
 \\
 single &       6420.3& 0.7&6410.2& 0.6   &6514.4&0.6& 6327.5&0.6 &6516.3&0.7\\
 narrow 1 &	6397.3&0.9&6403.5&0.8&6516.5&1.3&6325.4&1.4&6518.3&1.2\\
 narrow 2 &	6457.1 &0.9 &\ldots&\ldots &\ldots&\ldots&6333.2&1.3&\ldots&\ldots\\
 broad &	6418.4 &4.4 &6414.1&	1.8&	6511.5&4.8&\ldots&\ldots&6499.8&3.5\\
\\
 \hline
\end{tabular}
\end{center}
\end{table*}

\subsection{Relation between H$\alpha$ Luminosities and Velocity Dispersion} 
\label{sec:L_Sigma}

\begin{figure*}
\begin{center}
\caption[LvsSigma]{$\log(L) - \log(\sigma)$ relation for our HII regions. Luminosities and velocity dispersions are derived from our spectrophotometric data. The plot includes results from narrow components (labelled as \textbfit{n}) and single profile fits (labelled as \textbfit{s}). We add the data from \cite{2010MNRAS.406.1094F} (n, nA and nB for narrow, and g for single components) identified by color solid error bars and with numbers (from 1 to 6): NGC\,7479-I (1) in red, NGC\,7479-II (2) in green, NGC\,7479-III (3) in yellow, NGC\,6070-I (4) in maroon, NGC\,6070-II (5) in violet and NGC\,6070-IV (6) in magenta. A few Giant H{\sc ii} Regions from \cite{B02} (blue dashed error bars) together with their linear fit to their ``young'' Giant H{\sc ii} Regions are plotted as a reference value. The luminosities are not corrected for extinction.}
\label{Fig:elesigma2}
\vspace{1.0cm}
\includegraphics[angle=0,width=15cm,height=10cm]{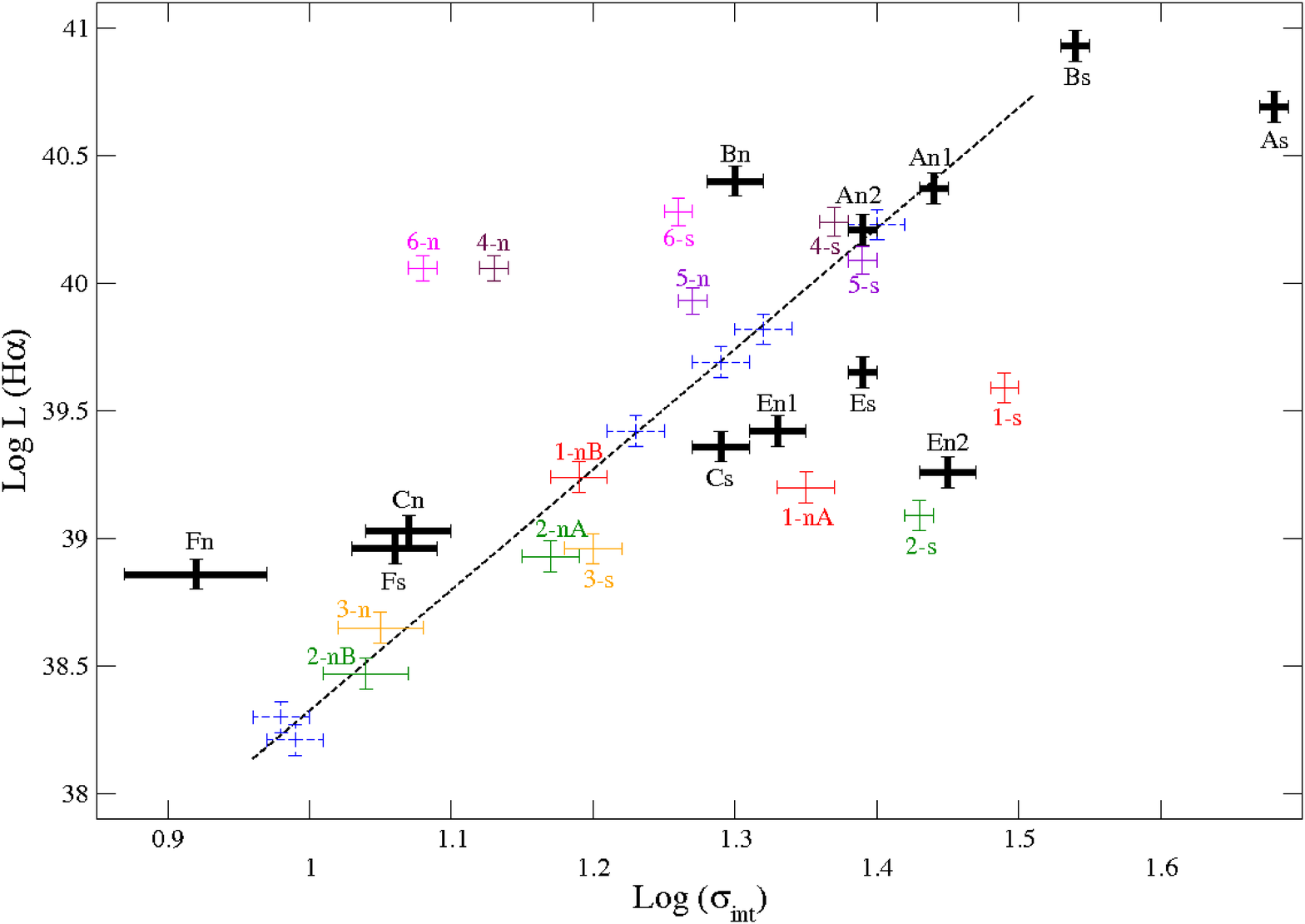}
\end{center}
\end{figure*}

Figure \ref{Fig:elesigma2} shows the distribution of \Haro\ knots in the L(H$\alpha$) vs. $\sigma$ plane. Each knot is labelled from A to F as used throughout this paper. Narrow components are identified with subscript \textbfit{n} while subscript \textbfit{s} refers to line widths measured by fitting a single Gaussian component to the line. Also plotted are data on NGC\,6070 and NGC\,7479 from \cite{2010MNRAS.406.1094F} and Giant \HII\ Regions from \cite{B02} together with the regression line fitted by those authors to their ``young'' Giant \HII\ Regions. H$\alpha$ luminosities were derived from the fluxes measured directly from the component fitting to our echelle spectra, and using distances as published by \cite{RC3.9} for \Haro. The correction for reddening could change flux values by a factor up to 1.72 (0.24 in logarithm) \citep{2009A&A...508..615L} but, as we cannot determine it for every regions, we prefer to list the uncorrected ones.
 
Similar to what was found by  \cite{2010MNRAS.406.1094F}, the location of the studied regions is far from random. \Haro\ \HII\ regions show a correlation between their luminosities an velocity dispersions. If we focus on the fit of single components they seem to follow the regression found for virialised systems. The outstanding exception to this trend  is region A, but this can be attributed to the fact that the profile can be split in two narrow components, resembling the behaviour found for regions NGC\,7479-I and NGC\,7479-II \citep{2010MNRAS.406.1094F}. The narrow components of the composite fits show, as expected, relatively smaller luminosities and velocity widths, but they still lie around the same linear regression. In this case, it is region F that fails to follow the trend but, as mentioned earlier, its low velocity dispersion makes it a very unlikely candidate for a Giant \HII\ Region.
It can be seen from the figure that while narrow components tend to cluster around the L(H$\alpha$) $\propto \sigma^{4}$ relation, single component fits would provide a flatter power-law exponent. This is probably due to the broad component contributing a substantial part of the total H$\alpha$ flux.
Although not included in the figure, the broad components are located, as expected, in a parallel sequence shifted towards lower luminosities and/or higher velocity dispersions.


\begin{table*}
\caption[Luminosities]{Intrinsic velocity dispersions ($\sigma_{int}$) and the H$\alpha$ luminosity (L(H$\alpha$)$_{s}$) for a fitted single Gaussian component, together with the their associated errors, expressed in \kms\ and \ergs, respectively. We add the H$\alpha$ luminosity (L(H$\alpha$)$_{overall}$) obtained from the overall H$\alpha$ flux, uncorrected for reddening, and its error, in \ergs.}
\label{tabsigmaL}
\begin{center}
\begin{tabular}{@{}cccc@{}}
\hline
knot &$\sigma_{int}$ & L(H$\alpha$)$_{s}$& L(H$\alpha$)$_{overall}$ \\
\hline
\Haro\,A & 47.3$\pm$0.9& 4.9$\pm$0.7\,$\times$\,10$^{40}$&4.9$\pm$0.7\,$\times$\,10$^{40}$\\
\Haro\,B&34.4$\pm$0.8& 8.6$\pm$1.2\,$\times$\,10$^{40}$& 8.9$\pm$1.3\,$\times$\,10$^{40}$\\
\Haro\,C&19.3$\pm$0.8&2.3$\pm$0.3\,$\times$\,10$^{39}$&2.4$\pm$0.3\,$\times$\,10$^{39}$\\
\Haro\,E& 24.4$\pm$0.9&4.5$\pm$0.6\,$\times$\,10$^{39}$&4.5$\pm$0.6\,$\times$\,10$^{39}$\\
\Haro\,F&  11.6$\pm$0.8&9.1$\pm$1.3\,$\times$\,10$^{38}$&9.5$\pm$1.3\,$\times$\,10$^{38}$\\
\\
 \hline
\end{tabular}
\end{center}
\end{table*}

 \section{Summary}

From high resolution spectroscopy of \Haro\ we have performed a thorough analysis of the emission line profiles of several knots considering multiple component fits to their profiles. Our results can be summarised as:
\begin{itemize}
\item Giant \HII\ Regions of \Haro\ show a complex structure within the profile of all their emission lines, detected both in recombination and forbidden lines.
\item The brightest emission lines can be split in at least two strong narrow components plus an underlying broad component.
\item Although regions tend to follow the galaxy kinematics, the narrow components of knot A have relative velocities that are too large to be explained by galactic rotation. This behaviour can be explained if we are observing the orbital motion around their gravity center.
\item Almost all knots follow the relation found between luminosities and velocity dispersions for virialised systems, either when considering single profile fitting or the strong narrow components in more complex fits. Among these, the single one shows a relatively flatter slope.
\end{itemize}

The presence of more that one component in the Gaussian fits to the emission line profiles, such as those analysed in this paper, has been discussed in several previous studies. H\"agele et al.,~(\citeyear{H07,2009MNRAS.396.2295H,Hagele+10}) showed fits that involved the existence of broad and narrow components for the emission lines of the ionised gas in circumnuclear star-forming regions. Furthermore, Firpo et al.,~(\citeyear{2010MNRAS.406.1094F}) were able to detect two distinct components within the narrow feature of the emission lines in the brightest Giant H{\sc ii} Regions NGC\,7479-I and II. Although we still lack the number of objets needed to support a generalization of this trend, these data show that the presence of multiple kinematical components among extragalactic star-forming regions cannot be overlooked. They might have a strong impact on subsequent analysis that rely on basic parameters, such as the velocity dispersion and chemical abundance of the ionised gas, the inferences about the nature and strength of the source of ionization, or the classification of the activity in the central regions of galaxies.

 It is necessary to map these regions with high spectral and spatial resolution and much better S/N ratio to disentangle the origin of these different components. 2D spectroscopy performed with Integral Field Unit (IFU) is the ideal tool to tackle this issue.

\section*{Acknowledgements}

We are grateful to the director and staff of LCO for technical assistance and warm hospitality.
We appreciate the comments and suggestions by the referee, \'Angel R. L\'opez-S\'anchez,  which significantly improved this paper.
This research has made use of the NASA/IPAC Extragalactic Database (NED) which is operated by the Jet Propulsion Laboratory, California Institute of Technology, under contract with the National Aeronautics and Space
Administration.  
Support from the Spanish \emph{Ministerio de Educaci\'on y Ciencia} (AYA2007-67965-C03-03, AYA2010-21887-C04-03), and partial support from the Comunidad de Madrid under grant S2009/ESP-1496 (ASTROMADRID) is acknowledged.
VF and GB would like to thank the hospitality of the Astrophysics Group of the UAM during the completion of this work.

\clearpage 

\label{biblio} 
\bibliography{biblos} 
\bibliographystyle{mn2e} 

\end{document}